\documentstyle[12pt,aaspp4]{article}
\def\plotone#1{\centerline{\epsfxsize=5.0in\epsfbox{#1}}}
\def\plotonesmall#1{\centerline{\epsfxsize=3.5in\epsfbox{#1}}}

\def\BE{\begin{equation}}
\def\BEL#1{\begin{equation}\label{#1}}
\def\EE{\end{equation}}
\newcommand{\COBE}{{\it COBE}}
\newcommand{\MAP}{{\it MAP}}
\newcommand{\ISO}{{\it ISO}}
\newcommand{\IRAS}{{\it IRAS}}
\newcommand{\HI}{H\,{\scriptsize I}}
\newcommand{\HII}{H\,{\scriptsize II}}
\newcommand{\Halpha}{H$\alpha$}
\newcommand{\etal}{{et al.}}

\newcommand{\nWpMMSr}{{\rm ~nW~m}^{-2}{\rm sr}^{-1}}
\newcommand{\ergcmHzsr}{{\rm ~erg~cm}^{-2}{\rm s}^{-1}
            {\rm Hz}^{-1}{\rm sr}^{-1}}

\newcommand{\dfdegree}{^\circ}
\newcommand{\degree}{^\circ}

\newcommand{\cm}{{\rm ~cm}}
\newcommand{\mm}{{\rm ~mm}}
\newcommand{\mJy}{{\rm ~mJy}}

\newcommand{\MJypSr}{{\rm ~MJy~sr^{-1}}}
\newcommand{\GHz}{{\rm ~GHz}}
\newcommand{\K}{{\rm ~K}}
\newcommand{\eV}{{\rm ~eV}}
\newcommand{\keV}{{\rm ~keV}}

\newcommand{\GeV}{{\rm ~GeV}}
\newcommand{\TeV}{{\rm ~TeV}}
\newcommand{\pc}{{\rm ~pc}}

\newcommand{\ombar}{\tilde{\omega}}  % perihelion

\begin{document}

\title{Detection of a Far IR Excess with DIRBE at 60 and 100
Microns}

%
% EDITOR:  Recommended running header is:
% Far IR Excess at 60 and 100 Microns
%

\author{Douglas P. Finkbeiner \& Marc Davis}
\affil{University of California at Berkeley, Departments of Physics and 
Astronomy, 601 Campbell Hall, Berkeley, CA 94720}
\authoremail{dfink@astro.berkeley.edu, marc@coma.berkeley.edu}
\and
\author{David J. Schlegel}
\affil{Princeton University, Department of Astrophysics,
Peyton Hall, Princeton, NJ 08544}
\authoremail{schlegel@astro.princeton.edu}

%------------------------------------------------------------------------------
% ABSTRACT
%------------------------------------------------------------------------------

\begin{abstract}
From analysis of the DIRBE weekly averaged sky maps, we have detected
substantial flux in the $60\micron$ and $100\micron$ channels in
excess of expected zodiacal and Galactic emission.  Two methods are
used to separate zodiacal light from more distant emission.  Method I
makes use of the time-dependence of the North-South annual variation
observed at the ecliptic poles.  This method is robust against errors
in the inter-planetary dust (IPD) model, but does not demonstrate
isotropy of the background.  Method II measures the ecliptic latitude
dependence of the dust over a range of ecliptic latitudes ($|\beta| >
35\degree$) at solar elongation $e=90\degree$.  This allows the excess
to be determined in each week of the DIRBE mission for high
redundancy, but the results depend weakly on the IPD model.  Both
methods give consistent results at $60\micron$ and $100\micron$.  The
observed signal is consistent with an isotropic
background at the level $\nu I_\nu = 28.1 \pm 1.8 \pm 7(\rm{syst})\nWpMMSr$ at
$60\micron$ and $ 24.6 \pm 2.5 \pm 8\nWpMMSr$ at $100\micron$. 

The IR excess detected at 140 and $240\micron$ by these methods agrees
with previous measurements, which are thought to be the cosmic
infra-red background (CIB).  The detections at 60 and $100\micron$ are
new.  The integrated IR excess in the window $45-125\micron$ is $23
\pm 8\nWpMMSr$, to be added to the $18 \pm 4\nWpMMSr$ previously
measured with the DIRBE and FIRAS instruments in the window $125 -
2\mm$.  While this new excess is not necessarily the CIB, we have
ruled out all known sources of emission in the solar system and
Galaxy.  We therefore tentatively interpret this signal as the CIB and
consider the implications of such energy production from the viewpoint
of star formation efficiency and black hole accretion efficiency.
However, the IR excess exceeds limits on the CIB derived from the
inferred opacity of the IGM to observed TeV photons, thus casting
doubt on this interpretation.  There is currently no satisfactory
explanation for the $60-100\micron$ excess.

\end{abstract}

%------------------------------------------------------------------------------
% SUBJECT HEADINGS
%------------------------------------------------------------------------------
\emph{Subject headings:}
cosmology: observations --- diffuse radiation --- infrared: ISM: continuum --- 
interplanetary medium --- galaxies: active --- gamma rays: observations

%------------------------------------------------------------------------------
% INTRODUCTION
%------------------------------------------------------------------------------
\section{INTRODUCTION}
\label{sec_intro}

The extra-galactic background light (EBL), from optical to sub-millimeter
wavelengths, records the 
energetics history of galaxy formation.  This background is the cosmic
relic of star formation, AGN, and black hole formation. 
The existence of such
a background was discussed first in the optical
and near IR (\cite{partridge67}) and then at other wavelengths
(\cite{low68}; \cite{peebles69}; \cite{harwit70}; \cite{kaufman76};
\cite{dube79}).  However, measurement of an unresolved background is
very difficult at most wavelengths because of numerous foregrounds
which may be many times brighter.  Direct measurement of the
individual background sources by deep, high resolution imaging has
only recently become possible, and only at selected wavelengths.

\subsection{Current Knowledge of Extragalactic Background}

The Infra-Red Astronomical Satellite (\IRAS) in 1983 obtained the
first full-sky census of far infrared (FIR) point sources.  Some
$300,000$ point sources, including $\sim 20,000$ galaxies, were
detected in four bands from $12-100\micron$ (see \cite{iras88}).
Optical follow-up indicated that typically 30\% of the bolometric
luminosity of these galaxies is radiated in the FIR (\cite{soifer91}),
presumably as thermal radiation from dust heated by optical/UV
radiation.  In the case of ultra-luminous IR galaxies (ULIRGs) up to
95\% of the bolometric luminosity is radiated in the FIR (e.g.,
Sanders \& Mirabel 1996), suggesting that optically obscured galaxies
might produce a substantial fraction of the extragalactic background
light.  Mid IR ($12-25\micron$) emission from most ULIRGs is centrally
concentrated, consistent with AGN activity (\cite{soifer99}).  

\begin{figure}[t]
\plotone{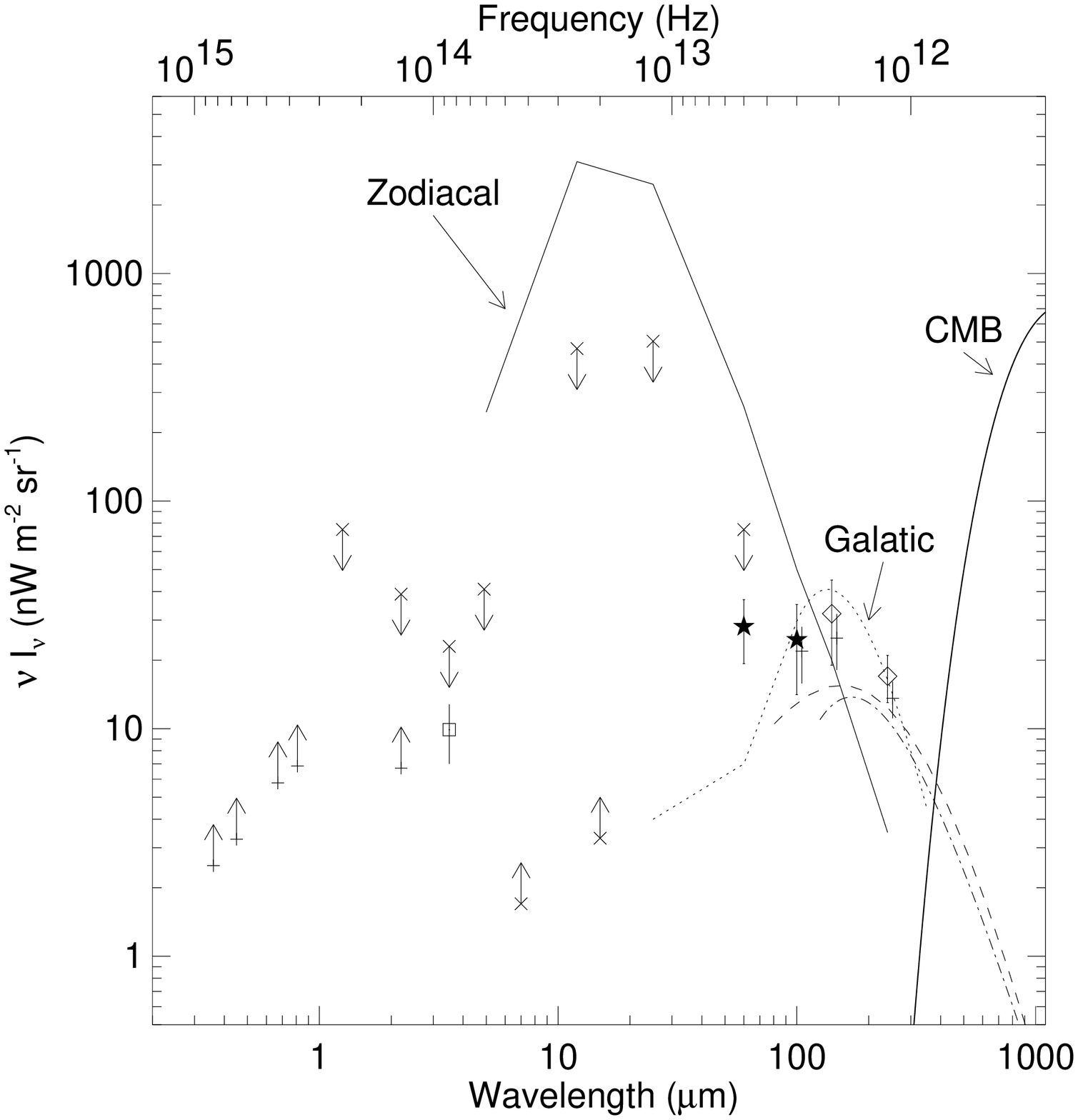}
\caption{Background radiation from UV to submillimeter wavelengths.
Lower limits (HST; \cite{pozzetti98}); square
(DIRBE; \cite{dwek98b}); lower limits (ISO; \cite{altieri99}); upper limits
(DIRBE; \cite{hauser98}); diamonds (DIRBE; \cite{sfd98}); crosses
(DIRBE; \cite{hauser98});
dash-dot line (FIRAS; \cite{fixsen98}).  
The filled stars are DIRBE measurements presented in this work.  
In all cases lower limits are derived from direct number counts,
while upper limits and measurements are obtained by
subtracting all known foregrounds from the observed sky surface
brightness. 
The dashed line is a simple model motivated
by Blain \etal (1999a).  The CMBR is shown as a thick line.
Typical high-latitude ISM brightness is shown by a dotted line, and
approximate IPD brightness at ecliptic poles is a thin line.
}
\label{fig_cibfig}
\end{figure}

More recently, Puget \etal\ (1996) used the \COBE\ Far InfraRed
Absolute Spectrophotometer (FIRAS) data to constrain the FIR/SMM
background at longer wavelengths.  They found that the integrated
energy of the EBL in the $200\micron - 2$mm window is comparable to
that emmited at optical/near IR wavelengths.  This picture was
confirmed by Guiderdoni \etal\ (1997) who concluded that the majority
of high-$z$ star formation may be hidden by dust.  The FIRAS
measurement was greatly refined by Fixsen \etal\ (1998) who obtained a
fit to the CIB of the form
\BE
I_\nu = (1.3\pm 0.4) \times 10^{-5} (\nu/\nu_0)^{0.64\pm0.12}
B_\nu(18.5\pm1.2\K)
\EE
in the interval $150 < \nu < 2400\GHz$ ($2000-125\micron$), where
$\nu_0=3000\GHz$ and $B_\nu(T)$ is the Planck function.  The
integrated intensity observed in this frequency interval is
$14 \pm 4\nWpMMSr$.  This curve is the dotted line shown in Figure
\ref{fig_cibfig}. 

Recent efforts to resolve this background into discrete sources with 
the SCUBA detector on the James Clerk Maxwell Telescope have been very
successful (\cite{smail97}; \cite{hughes98}; \cite{barger98};
\cite{eales99}).  By extending the counts to a limit of 0.5 mJy in
cluster-lensed fields, enough counts are found to account for
most of the expected $850\micron$ background of $\nu I_\nu = 5\pm 2
\times 10^{-1} \nWpMMSr$ (\cite{blain99b}).

At shorter wavelengths, the background is not resolvable with current
instruments, and zodiacal emission from interplanetary dust (IPD)
hampers detection of the unresolved background.  The zodiacal emission
peaks at $\sim 25\micron$ and dominates any expected signal from the
CIB in most DIRBE channels.  Nevertheless, the background has been
measured at $140\micron$ and $240\micron$ by Schlegel,
Finkbeiner, \& Davis (1998).  A more thorough analysis by the DIRBE
team (\cite{hauser98}) improved those results, and also provided upper
limits in the other 8 DIRBE channels (see Figure \ref{fig_cibfig}).
The FIRBACK survey at $175\micron$ (\cite{puget99}) using the
\emph{Infrared Space Observatory} (\ISO) detected point sources to a
flux limit of $200\mJy$, yielding an integrated flux significantly
higher than a simple extrapolation of IRAS counts would have
predicted, but still much lower than the observed background.  Puget \etal\
(1999) claim that a plausible extrapolation of the counts down to
$10\mJy$ would account for the entire background at $175\micron$.

At wavelengths $\lambda \le 3.5\micron$, emission from Galactic stars
dominates that from zodiacal dust.  Using ground-based $2.2\micron$
counts to remove the stellar foreground, and adopting a value of
$7.4\nWpMMSr$ at $2.2\micron$, Dwek \& Arendt (1998) measured the
background at $3.5\micron$ to be $\nu I_\nu \sim 10\pm 3
\nWpMMSr$. 
Gorjian, Wright, \& Chary (1999) make different assumptions about the
stellar foregrounds, yet arrive at a similar result: $16.2\pm6.4
\nWpMMSr$ at $2.2 \micron$ and $9.3\pm 3\nWpMMSr$ at $3.5 \micron$. 

At optical wavelengths the Hubble Space Telescope can resolve a large
fraction of the extragalactic background.  By integrating the light of
the resolved galaxies in the HDF, Pozzetti \etal\ (1998) find that the
brightness of the extragalactic sky is $2.1^{+0.4}_{-0.3}\times
10^{-20}\ergcmHzsr$ in I-band ($\sim 0.8\micron$), which is equivalent
to $\nu I_\nu = 8 \nWpMMSr$.  In the other Hubble broadband filters,
this background obeys $\nu I_\nu \sim \nu^{-1}$ for 2000 - 8000\AA.
The HDF counts in each filter appear to flatten at the faint end,
indicating a possible convergence, although other authors find hints
of an optical/UV background as much as twice this large
(\cite{bernstein97}).  Comparing the latest FIR background with the
optical/UV background has supported the view that $\sim 2/3$ of the
starlight in high-$z$ galaxies is reprocessed by dust into FIR
radiation, a much higher percentage than in local galaxies.  Where,
then, are the sources of the FIR background?

It is well established that the ULIRGs are typically interacting
galaxies (see Sanders \& Mirabel 1996), and Williams \etal\ (1996) find
that most high-$z$ objects in the HDF are interacting.  What is not
very well established is the dominant energy source in these objects,
especially at the high-luminosity end.  Certainly, many of these
objects are powered by starbursts; only the very brightest objects so
far observed are primarily powered by AGN (\cite{genzel98,lutz98}).
But this scenario may change at high redshift.  Measurements of the 
X-ray background provide some constraint on AGN activity, unless the
very brightest AGN are so obscured that even $10\keV$ photons cannot
escape. 

\subsection{Interpretation}
The above observational evidence allows for a coherent interpretation.
Interacting ULIRGs at high-$z$ undergo violent episodes of dusty star
formation, with AGN as a minor contributor to the energy.  Roughly 2/3
of this energy is reprocessed to the FIR.  Accounting for present day
metallicity of galaxies and the IGM enrichment, the energy from stars
allows for an integrated extragalactic background light (EBL) of
$I_{\rm bol} = 50 / (1+z_f) \nWpMMSr$ (see
\S \ref{sec_crisis}), which is roughly consistent with 
current measurements if the formation redshift $z_f \la 2$.

The IR excess measurements presented in this paper, if they are
interpreted as an extragalactic background, would push this paradigm
to the limit.  They would indicate an integrated EBL flux in the far
IR of $\sim 40 \nWpMMSr$ with a hot spectrum, possibly suggesting that
AGN dominate the energy input in early galaxies, at least at short
wavelengths.  This interpretation might also imply the presence of up
to $\sim 0.15\%$ of all baryons in black holes, and possibly violate
constraints on the X-ray background.  Furthermore, recent measurements
of high energy gamma rays place limits on the opacity of the IGM,
which is primarily due to pair production on CIB photons.  
The interpretation of the $60-100\micron$ excess as CIB raises serious
problems; however, we have been unable to identify any
alternative source in the Galaxy, local bubble, or Solar system that
can account for the emission.

\subsection{Organization of Paper}
We review previous foreground models in \S\ref{sec_foregrounds}, and
present our two procedures for extracting the IR excess signal in \S
\ref{sec_methods}.  Our assessment of various systematic errors is
given in \S 4.  In \S 5 we discuss the energy crisis resulting if
this emission is interpreted as an extragalactic background.  In
addition, we discuss the constraint on the density of CIB photons
implied by the observation of TeV photons from nearby AGN.  These
difficulties, as well as general conclusions and prospects
for the future appear in \S 6.

%------------------------------------------------------------------------------
% FOREGROUNDS
%------------------------------------------------------------------------------
\section{FOREGROUNDS}
\label{sec_foregrounds}
\subsection{Zodiacal Emission}

The main difficulty in measuring the CIB at $60-240\micron$ is
contamination from zodiacal light, which is thermal emission from the
interplanetary dust (IPD).  This emission is brightest in the DIRBE 12
and $25\micron$ bands, falling approximately as a blackbody at longer
wavelengths.  The emission at the ecliptic poles is $\nu I_\nu\sim 260$ and 
$50\nWpMMSr$ in the DIRBE 60 and $100\micron$ bands respectively
(Fig. \ref{fig_cibfig}).
Emission in the ecliptic plane is $\sim 3-4$ times brighter. 
This emission must be carefully removed in order to measure the much
fainter extragalactic background. 

The IPD cloud is difficult to model, especially at low ecliptic
latitude.  At high latitude, one looks through dust in the
neighborhood of the Earth, but at low latitude, the situation is more
complex: the dust density and temperature vary significantly along a
line of sight, several distinct dust rings are seen, and a density
resonance in the Earth's orbit is observed.  These factors make it
nearly impossible to model the zodiacal light at low latitudes from
$5-100\micron$.  Furthermore, temperature and density variations
appear even in the near-Earth dust, beyond the expected variation due
to changes in the Earth-Sun distance through the year.

\subsubsection{SFD98 Zodiacal Emission Model}

A simple approach to the problem of separation of zodiacal light from
other emission is presented in Schlegel, Finkbeiner, \& Davis (1998; hereafter
SFD98).  They use the  $25\micron$ map as a spatial template of the
zodiacal light.  At high ecliptic latitude, most of the dust is less
than 0.4 AU from Earth and has a fairly uniform temperature.  Near the
ecliptic plane, one sees dust of varying temperature out to several
AU.  Therefore, this $25\micron$ template does not extrapolate to
longer wavelengths in a linear way, and another level of detail must
be added.

To good approximation, the error made by a linear model is a function
only of ecliptic latitude.  Therefore, one must modulate the
$25\micron$ map by some reasonable function of ecliptic latitude.
Rather than choose standard basis functions such as a series of
($m=0$) spherical harmonics, SFD98 instead employs a function
determined by the dust itself - simply the $25\micron$ flux binned in
ecliptic latitude.  This modulation is adequate with only one term in
the expansion, and comprises the so-called ``quadratic zodiacal light
model'' (see SFD98, equation 3).

The success of such a method depends on a few assumptions.  One is
that the Galactic dust cirrus is negligible at $25\micron$.  SFD98
showed that, in any place where the $100\micron$ flux is less than
$100\MJypSr$, the $25\micron$ flux due to cirrus causes at most a 1
percent error in the final map.  Another assumption is that there is
no CIB in the $25\micron$ map.  This approach will not be sensitive to
any CIB component that has the same spectral shape as the zodiacal
emission.  Current models of the CIB predict no substantial flux at
$25\micron$, and certainly less than the flux at $140\micron$ (in $\nu
I_\nu$ units).  If the CIB flux at $25\micron$ were as high as it is
at $140\micron$ ($25 \nWpMMSr$), the offset introduced in the 60 and
$100\micron$ cirrus maps would be negligible ($3\nWpMMSr$ at $60\micron$
and $1\nWpMMSr$ at $100\micron$). 

This ``quadratic model'' was used by SFD98 to separate cirrus
(Galactic ISM) emission from CIB and zodiacal light.  At 140 and
$240\micron$, the zodiacal emission is faint enough that its latitude
dependence provides enough information to separate it from the CIB; at
shorter wavelengths, more sophisticated methods are required. 

\subsubsection{Goddard Zodiacal Emission Model}

The model invented by the Goddard team (\cite{kelsall98}) is an
ambitious attempt to parameterize the full spatial-temporal dependence
of the IPD emission.  It contains six components, a smooth cloud,
circumsolar ring, a density enhancement following the Earth and in
resonant lock, and also three dust bands near the ecliptic at 3 AU.
Certainly, the complexity of the dust cloud justifies the 46 model
parameters, and a sophisticated model is required to establish the
isotropy of the background.  However, even though the parameterization
is physically well motivated, a simple subtraction of the model from
the data leaves significant unexplained residuals at $\lambda \le
100\micron$.  These residuals may result from detector gain drifts,
or small variations in dust density and temperature.  For this reason,
a more robust approach is required. 

The two methods given in \S \ref{sec_methods} each construct a
dimensionless parameter that is robust (at first order) with respect to
gain drifts and dust density and temperature variation.  This
parameter is evaluated for each of the 40 weeks of the DIRBE mission,
using the DIRBE weekly skymaps with Galactic emission subtracted.
Before discussing these methods in detail, we next consider the
Galactic (ISM) emission removal from each weekly skymap.

\subsection{Galactic ISM dust emission}
Emission from Galactic dust in the ISM (cirrus) is typically
comparable in brightness to the IR excess at $100\micron$, and about
1/4 as bright at $60\micron$ (Fig. \ref{fig_cibfig}).  As pointed out
in the ``zodiacal light'' fits in SFD98, it is easier to separate the
Galactic cirrus from the sum of the other emission components than it
is to isolate the CIB.  SFD98  describe FIR maps that have a
``quadratic zodiacal light model'' (actually zodiacal light plus CIB)
removed, which we call the ``cirrus-correlated maps.''  These maps
are constructed by removing a quadratic zodiacal emission model from
the DIRBE annual average maps so as to maximize the correlation of 
the cirrus map with the Leiden-Dwingeloo \HI\ survey at high galactic
latitude (see SDF98 for details).  Because only 3 degrees of freedom
are used to fit the HI data to the full-sky DIRBE maps,
the structure of the Galactic ISM
is derived directly from the DIRBE
maps, not from \HI\ data.  These annual-average
DIRBE maps are subtracted
from each weekly average DIRBE map before the processing described in
\S \ref{sec_methods} is performed.  The structure of the Galactic
cirrus and extragalactic point sources are therefore almost perfectly
subtracted, with remaining residuals attributed to detector gain drift
and shifted effective pixel centers from week to week.

The uncertainty in our final IR excess measurements resulting from the
cirrus emission is discussed in \S \ref{sec_wim}.  
There may also be substantial emission from dust in the warm ionized
medium (WIM), as traced by \Halpha\ and pulsar dispersion measures.
However, the WIM is well correlated with the CNM on spatial scales of
interest, and this contribution should not result in a significant
error.  Section \ref{sec_wim} gives a complete discussion of
systematic uncertainty associated with the WIM dust. 

The next section addresses the
brightest foreground, the zodiacal light.

%------------------------------------------------------------------------------
% ZODIACAL EMISSION REMOVAL:  TWO METHODS
%------------------------------------------------------------------------------
\section{ZODIACAL EMISSION REMOVAL:  TWO METHODS}
\label{sec_methods}

\subsection{Method I}

\begin{figure}[t]
\plotone{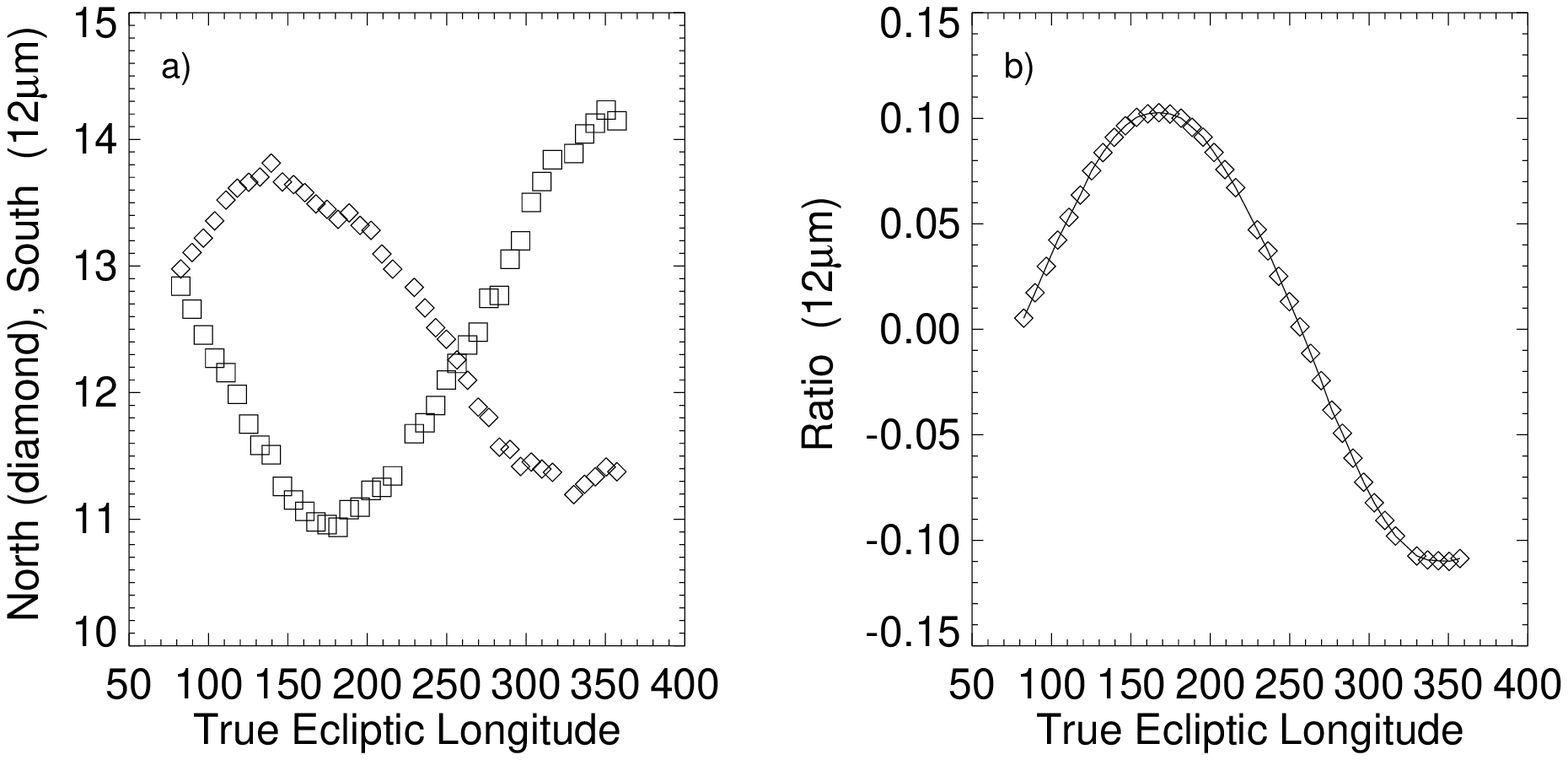}
\caption{Weekly $12\micron$ DIRBE data at the ecliptic poles.
(a) Weekly flux at the north (diamonds) and south (squares) ecliptic poles in
$\MJypSr$, plotted as a function of the true heliocentric ecliptic longitude of
Earth. 
(b) The dimensionless ratio of $(N-S)/(N+S)$ defined in eq. (\ref{equ_rb}).
}
\label{fig_r12}
\end{figure}

\begin{figure}[t]
\plotone{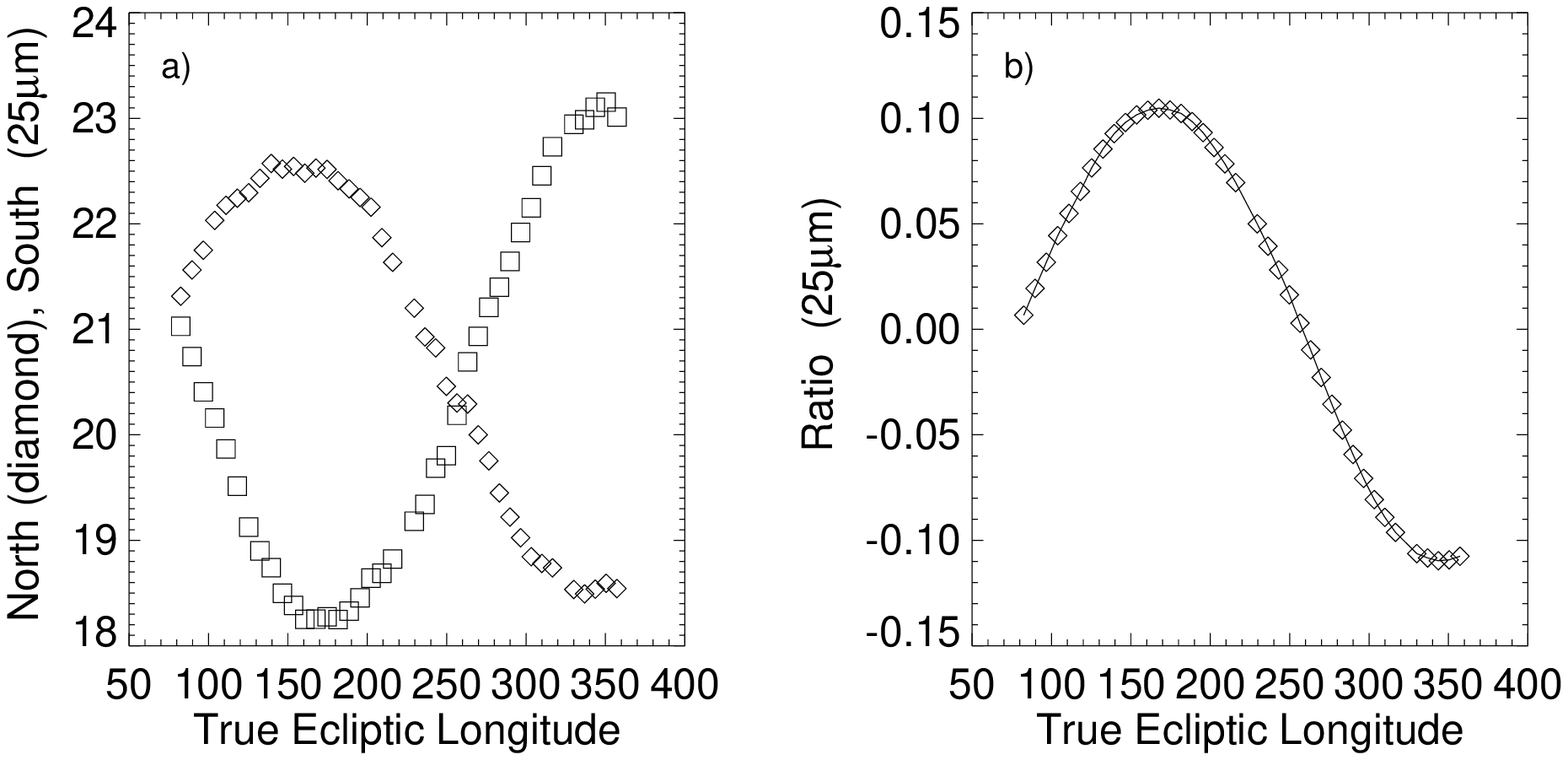}
\caption{Weekly $25\micron$ DIRBE data at the ecliptic poles.
(a) Weekly flux at the north (diamonds) and south (squares) ecliptic poles in
$\MJypSr$, plotted as a function of the true heliocentric ecliptic longitude of
Earth. 
(b) The ratio of $(N-S)/(N+S)$ defined in eq. (\ref{equ_rb}).
}
\label{fig_r25}
\end{figure}

Our first method makes use of the north-south
variation of the zodiacal emission observed by DIRBE as a function of
time.  The Interplanetary Dust (IPD) cloud is inclined $\sim 2\degree$
with respect to the ecliptic, resulting in a north-south
asymmetry with a one year period (see Figs.
\ref{fig_r12} and \ref{fig_r25}).  At first, it appears that this
temporal variation is an undesirable complication in an already
complex problem.  However, this temporal variation allows us to probe
the dust cloud in the $z$ direction (normal to the dust plane) and
separate zodiacal emission from the other components.  Because the
temperature and density of the IPD change throughout the year, it is
useful to consider a dimensionless parameter that corresponds to our
$z$ position in the IPD cloud, a number for which uncertainties in the
overall IPD density and temperature cancel out.  Therefore, it is
convenient to define a dimensionless ratio,
\BEL{equ_rb}
R_b\equiv \frac{N_b-S_b}{N_b+S_b},
\EE
where $N_b$ ($S_b$) is the total DIRBE flux at the north (south)
ecliptic pole in band $b$.  In bands where the zodiacal light
overwhelms emission from cirrus, CIB, or other contamination, this
quantity is related only to the $z$ position of Earth in the dust
cloud - not to density or temperature.  Because variations in dust
temperature, density, and detector gain drifts cancel out in $R_b$, it
is far more robust than any absolute measurements.  In fact, $R_b$ is
nearly independent of IPD model. We are assuming that the
$z$-dependence of IPD emission per volume has the same functional form
for each waveband, which should be true for dust near the Earth.  This
requirement is not strictly satisfied by the IPD, but is justified in
section \ref{sec_errors}.

Because it is exceedingly difficult to model the IPD cloud along all
lines of sight at all times, we restrict ourselves to analysis of
patches within $5\degree$ of the ecliptic poles.  These regions contain
data for every week of the DIRBE mission, and effects dependent on
solar elongation angle cancel out to first order.  We have prepared a mask
for pixels in this region, excluding those pixels not present in symmetric
combinations for each week.  Details are given in \S\ref{sec_err_results}.

Our resulting measurements of the IR excess at 60 and $100\micron$ depend
upon two main assumptions:  (1) that the amplitude of the annual
variation in $R$ would be nearly constant in all wavebands in the
absence of the excess, and (2) that our $100\micron$ cirrus map (SFD98) is
correctly zero-pointed.  Both of these assumptions have been explored
thoroughly, and are investigated in detail in \S\ref{sec_errors}. 

\begin{figure}[t]
\plotonesmall{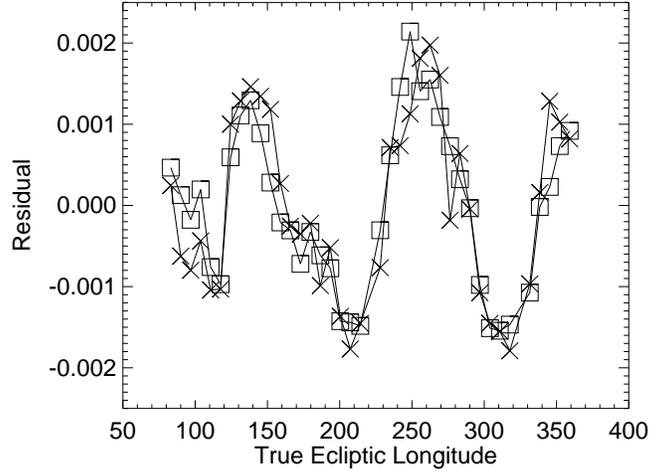}
\caption{
Model Residuals difference for 12(diamonds) and
$25\micron$.  A significant correlation is present in the residuals,
but at a level of approximately one part in 1000 of the total signal.
The interplanetary dust cloud may not be any smoother than this. 
}
\label{fig_zodyres}
\end{figure}

Plots of $R_{12}$ and $R_{25}$ are shown in Figures \ref{fig_r12}b and
\ref{fig_r25}b.  The
line is a simple sinusoidal model - the best fit for
\BE R_{model}=A\sin(\nu-\nu_0)+C \EE 
where $A$ is the amplitude, $\nu$ is the true anomaly (i.e. the angle,
measured at the Sun, between perihelion and the Earth $\approx$ true
longitude minus $102.8\dfdegree$), $\nu_0$ is a phase angle, and $C$
is a small constant.  The true anomaly, $\nu$, is used instead of mean
heliocentric longitude because it corresponds more directly to Earth's
$z$ position in the dust cloud.  The deviations between the model and
data are too small to see in Figures \ref{fig_r12} and
\ref{fig_r25}, and therefore are plotted in  Figure
\ref{fig_zodyres}.  The RMS dispersions of $R_{12}$ and $R_{25}$
relative to this simple model are 1\% of the amplitude of the sine
wave - and 0.1\% of the total emission at the poles.  Furthermore,
there is a strong correlation between the residuals at 12 and
$25\micron$ - suggesting that the residuals are physical.  This is a
powerful test of the relative photometric stability of the DIRBE 12
and $25\micron$ detectors, and allows us to proceed to the next step.

\subsubsection{Scale-Height of the Dust}

A common \emph{Ansatz} for $z$-dependence of density in a disk is
$\exp(-|z|/h)$ where $z$ is the height above the dust plane and
$h$ is some scale height.  Such a model has a density cusp at $z=0$,
but because we travel only $\sim 0.03$AU in the $z$ direction, the
effect of the cusp is negligible.  Another reasonable guess might be a
gaussian
\BE \rho = \rho_0 \exp(-z^2/2\sigma_z^2). 
\EE
This distribution has the convenient property that $\rho(r,z)$ is
smooth near $z = 0$, consistent with our sinusoidal fit.
A third model is the Goddard widened fan model (\cite{kelsall98},
Eq. [7]), with a vertical profile
\BE
f(\zeta)=\exp(-\beta g^\gamma)
\EE
where $\zeta \equiv |Z_c/R_c|$,
\BE
g= \left\{ \begin{array}{lr}
  \zeta^2/2\mu & {\rm ~for~} \zeta < \mu \\
  \zeta-\mu/2 & {\rm ~for~} \zeta > \mu    \end{array} \right.
\EE
and $\beta=4.14\pm0.067$, $\gamma=0.942\pm0.025$, and
$\mu=0.189\pm0.014$ are the best-fit values of the parameters.  

Results presented in this paper are derived using the gaussian model
and are indistinguishable from the other models.  All we require is
that the density distribution be reasonably smooth near $z=0$ and not
extend too high above the plane. 

How much north-south variation is expected?  Symmetry considerations
suggest that the dust plane is approximately aligned with the
invariable plane of the solar system, which is perpendicular to the
total angular momentum vector of the solar system.  The north pole of
this plane is at $\alpha_{2000}=273.85$, $\delta_{2000}=66.99$ or
$\lambda=17.8$, $\beta=88.42$ in ecliptic coordinates.  (We denote the
heliocentric mean ecliptic longitude of Earth with $L$ hereafter in
this paper, and reserve the usual symbol $\lambda$ for wavelength.)
This difference between the ecliptic plane and invariable plane is
critical for our method, but the actual inclination angle is nearly
degenerate with the dust cloud height in our fits.  Our results are
the same whether the $1.58\degree$ angle to the invariable plane, or the
$2.03\degree$ inclination of the
\cite{kelsall98} smooth cloud is used. 

The perihelion (minimum Earth-Sun distance) is at $L \approx
102.8\dfdegree$.  The ascending node of the putative dust plane is at
$107.8\dfdegree$.  Because they are near each other, the time of
maximal excursion from the dust midplane occurs when the Sun-Earth
distance is approximately 1 AU.  Therefore, the extreme values of $z$
are $\pm0.0276$ AU.  Within this range (according to the fit in Figure
\ref{fig_r25}) resides $\sim 10.4$\% of the dust.  Simple algebra
then gives a scale height of $\sigma_z=0.20 \pm 0.01$ AU for the
gaussian.  This is very similar to the FWHM of the Kelsall model.  Of
course, the density profile need not be a gaussian in $z$; if the
emissivity of the IPD can be written as $I_b(z,r)n(z,r)$ then we have
the model-independent constraint that

\BE \int_{low}^{hi} I_b(z,r)n(z,r) dz = 
0.104 \int_{-\infty}^\infty I_b(z,r)n(z,r) dz 
\EE

We only use this to justify our assumption that the
dust along a line of sight is near the Earth and of uniform
temperature.  None of the conclusions in this paper depend on the
functional form of the $z$-dependence.  

\subsubsection{Model Fit} 

Now that we have established a reasonable model for the $z$ dependence
of dust emission, let us formalize it a bit.  Suppressing the $b$
subscript for convenience, let us define N (S) to be the total
emission observed near the north (south) pole of the invariable plane:
\BE N \equiv Z_N + B_N \EE
\BE S \equiv Z_S + B_S \EE
 where $Z_N$ ($Z_S$) is the zodiacal light in
the north (south) and $B_N$ ($B_S$) is the time-independent background
in the north (south), including cirrus, CIB, Reynold's Layer, and any
other unknown backgrounds such as halo dust.  Let us also define
\BE Z \equiv Z_N + Z_S ; B \equiv B_N+B_S \EE
where $Z$ is the total column emission through the IPD plane,
while $B$ is \emph{twice} the average background. 

For simplicity, we assume that the emission per volume is constant
near the ecliptic plane, such that $R$ depends only on
$z=r\sin\theta\sin i$, with $r$ and $\theta$ suitably defined for an
elliptical orbit.  The position of Earth is parameterized by the true
anomaly, $\nu$, given in the almanac as
(\cite{almanac91}, p. E4):
\BE \nu = M + (2e-e^3/4)\sin M + (5e^2/4)\sin2M + (13e^3/12) \sin3M +
O(e^4) \EE
where $e$ is the eccentricity of Earth's orbit ($e \approx 0.0167$),
$M=L-\ombar$, $L$ is the mean longitude and $\ombar
\approx 102.8\dfdegree$ is the mean longitude of perihelion.
The true Earth-Sun distance is then 
\BE r=(1-e^2)/(1 + e\cos\nu) \EE
where $r$ is in AU.  We let $\Omega$ denote the nominal longitude of the
ascending node of the dust plane (approximately $L=77\dfdegree$) and
define $\theta$ as 
\BE \theta = \nu + \ombar - \Omega \EE
The height of Earth above the dust plane is given by 
\BE z = r\sin\theta \sin i \EE
where $i$ is the inclination of the dust plane ($i \approx 1.58\dfdegree$). 
The zodiacal emission at the ecliptic poles is given by
\BE Z_N = {Z\over 2} \left(1+ Ar\sin\theta\right) \EE
\BE Z_S = {Z\over 2} \left(1- Ar\sin\theta\right) \EE
and the ratio $R$ by 
\BE R = A'r\sin\theta + C. \EE
with the definitions 
\BE A' \equiv A\left({Z\over Z+B}\right) ; C \equiv {B_N-B_S \over Z+B}\EE
Here the constant factor $\sin i$ is absorbed in the $A$ coefficient
to emphasize that $A$ and $\sin i$ are degenerate parameters in this
model.  The results of this paper do not depend on
the value of $\sin i$ in detail, only that it be small and constant.
Physically, $A$ is the amplitude of the annual variation in $R$ due to
the zodiacal emission and $A'$ is the observed amplitude.

The total background may now be expressed in terms of the observables, $N$
and $S$:
\BE B = (N+S)\left(1-{A' \over A}\right), \EE 
and the difference, 
\BE B_N-B_S = C(N+S), \EE
which leads immediately to expressions for the north and south:
\BE B_N = {B + C(N+S) \over 2} \EE
\BE B_S = {B - C(N+S) \over 2} \EE

\subsubsection{Results}

Results of the fits in the DIRBE 5-$240\micron$ bands are presented in
Table \ref{table_cib_fit}.  The best-fit model parameters $A'$,
$\nu_0$, and $C$ are given for each waveband, as well as the derived parameters $Z$,
$B_N$, and $B_S$.  The value of $\nu_0$ is determined from the 12 and
$25\micron$ bands and adopted for the others.  Values of $A(\lambda)$ must be
assumed in order to derive $B$ from $A'$, and because the $25\micron$
band is dominated by zodiacal emission, $A\approx A'$. 
As can be seen from the
large errors for the 140 and $240\micron$ bands, this method breaks
down at long wavelengths where the zodiacal emission is weaker and the
S/N of the detectors is much lower.  Although the actual CIB level at
12 and $25\micron$ is unknown, it must be non-negative.  The
assumption of a significant CIB in these wavebands would only increase
the amount deduced for the longer wavelengths.  Therefore, we assume
the CIB at $25\micron$ to be zero to make a conservative
assessment of the emission at $60-240\micron$.  The systematic errors
introduced by this assumption are determined by computing the change
in the measured excess if the $25\micron$ CIB has the same $\nu I_\nu$ as
$140\micron$.  The large excess at $12\micron$ is very
uncertain, because of the larger model dependence at short
wavelengths.  The excess at $5\micron$ is sufficiently model dependent
that it should not be taken seriously. 

Values for $B_N$ and $B_S$ are shown separately in Table
\ref{table_cib_fit} to demonstrate that the $N-S$ difference due to
cirrus has been adequately removed from the weekly maps.  We take the
$A$ dependence calculated for the Goddard widened-fan model, but
consider other models in \S\ref{sec_errors} to estimate systematic
errors.  This slight model dependence will propagate into our final
systematic errors, but for now we retain the assumption that $A$ is
constant in every DIRBE band.

\subsubsection{Model Refinements}

The residual seen in Figure \ref{fig_zodyres} suggests a residual with
a 1/3 yr period at both 12 and $25\micron$.  We have added a few
parameters to the model to allow for two dust disks of different
thicknesses, inclinations, and ascending nodes (similar to the
``circumsolar ring'' in the Goddard model), and this improvement
removes the 1/3 yr period signal.  Another improvement was made to
account for possible emission from interstellar dust grains focussed
by the Sun into a cone in the downstream direction of the Sun's motion
relative to the local ISM.  Although inclusion of these effects
improves the $\chi^2$ of the fit, the change is not substantial, and
no significant variation in the derived IR excess results.  They are
therefore not considered further in this paper.

%Bill Reach says   T=261.5K +-1.5 blackbody - based on 5-16.5 mu
%measurements with ISO.  Some sign of an 11mu feature. 
%blackbody only good to ~ 15 percent. 
%This overpredicts zody at 60mu by 50% though. 
%Reach, W. T., \etal\ 1996  Astronomy and Astrophysics, v.315, p.L381-L384

\subsection{Method II}
\label{sec_method2}

In this section, we present a method to disentangle the CIB from
zodiacal emission by making use of the ecliptic latitude dependence of
the latter.  We define a dimensionless statistic, $\Xi$ that can be
measured independently for each of the 41 weeks of the DIRBE mission.
This quantity is designed to be insensitive to the zodiacal emission
model parameters.  At the same time, this statistic provides a
sensitive measure of any isotropic background.  Although this
statistic involves less robust assumptions about the IPD than method
I, we test its dependence on the parameters of the Kelsall \etal\
(1998) model.  The $\Xi$ statistic allows for a wider range of
statistical tests and addresses questions about isotropy.
Furthermore, the assumption of a small CIB value at $25\micron$ is not
required.

%\subsubsection{Definition of $\Xi$ }

In SFD98, the CIB was measured at $140\micron$ and $240\micron$ by
fitting and removing a $\csc|\beta|$ slab component to the
annual-average DIRBE skymaps at high Galactic latitude.  Of course,
the annual-average maps combine data from many solar elongation angles
averaged over 41 weeks (not 1 yr) and contain the resulting artifacts.
Moreover, there are theoretical and observational reasons to suspect
that the IPD is a ``modified fan'' and not a slab at all (see
\cite{kelsall98}, \S 4.2 for discussion).  In the limit where the
zodiacal emission is small compared to the CIB, this method gives
reasonable results, but fails badly at $\lambda \le 100\micron$.  The
basic idea can be used successfully, however.

\begin{figure}[t]
\plotonesmall{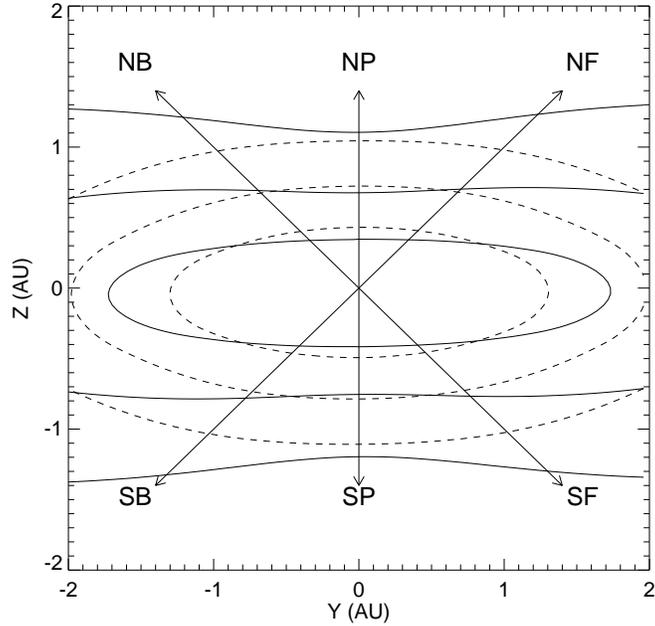}
\caption{Emissivity contours for $12\micron$ (dashed) and
$240\micron$ (solid).  The lines of sight used to define $\Xi$ in
Eq. (\ref{equ_xidef}) are labeled.  Earth is located in the middle of the
plot, and the Earth-Sun line goes into the page.  From this diagram,
it is clear that $\Xi$ is larger for longer wavelengths. }
\label{fig_emiscon}
\end{figure}

By working only with solar elongation $90\degree$ data in each weekly
map, the problem is conceptually simpler.  In Figure \ref{fig_emiscon}
we show the volume emissivity density contours of the $e=90\degree$
plane, i.e., the plane containing Earth and perpendicular to the
Earth-Sun line.  The axes are labeled with Cartesian ecliptic
coordinates, in AU.  The Earth is in the middle (at a time of year
when $x=1$, $y=z=0$), with lines of sight to the ecliptic poles
labeled NP and SP, and the ``forward'' direction of Earth's orbit to
the right.  Four other lines are drawn, all of which are at latitude
$|\beta| = 45\degree$.  They are labeled NF
for ``North-Forward,'' NB for ``North-Backward'' and so on.  A
convenient dimensionless ratio to define is:
\BEL{equ_xidef}
\Xi(\beta) \equiv \frac{NF+NB+SF+SB}{2 (NP+SP)}
\EE
This quantity is almost completely insensitive to vertical position in
the dust cloud, to a small inclination of the dust cloud with respect
to the ecliptic, dust temperature, or to nearly any other parameter in
the Goddard (Kelsall \etal\ 1998) model.  In fact, when $\Xi(\beta)$
is computed for the Goddard model (the most realistic model to date),
its annual variation is negligible ($< 0.001$) for the latitudes of
interest ($35\degree < \beta < 50\degree$).

If the zodiacal emission were approximately a slab, one would expect
the functional dependence $\Xi(\beta) = \csc(\beta)$.  It is notable,
however, that the $\Xi$ ``measured'' from the Goddard model at long
wavelengths is significantly greater than $\csc|\beta|$ in the range
of interest ($35\degree < \beta < 50\degree$).  This is because the
Rayleigh-Jeans emissivity density contours shown in Figure
\ref{fig_emiscon} (\emph{solid contours}) follow the volume density
contours closely, and the ``fan'' nature of the model causes an upward
curvature of the contours, resulting in $\Xi(\beta) >\csc|\beta|$.

At short wavelengths, the situation is reversed.  The emissivity is so
temperature sensitive that distance from the Sun is the overriding
concern.  In this case, the emissivity contours curve downward,
yielding $\Xi(\beta) < \csc|\beta|$ (Figure \ref{fig_emiscon},
\emph{dashed contours}). 

These deviations from the geometry of a uniform slab are fine points,
and do not affect the measurements at $60$ and $100\micron$, as we
shall see, but one must account for them carefully in
order to measure the background at $3.5-12\micron$, which is beyond
the scope of the current paper. 

For simplicity, we now return to the approximation that
$\Xi_0=\csc|\beta|$. 
Now let us consider the effect of an isotropic background, $B$, on
observed values of $\Xi$.

\begin{figure}[t]
\plotone{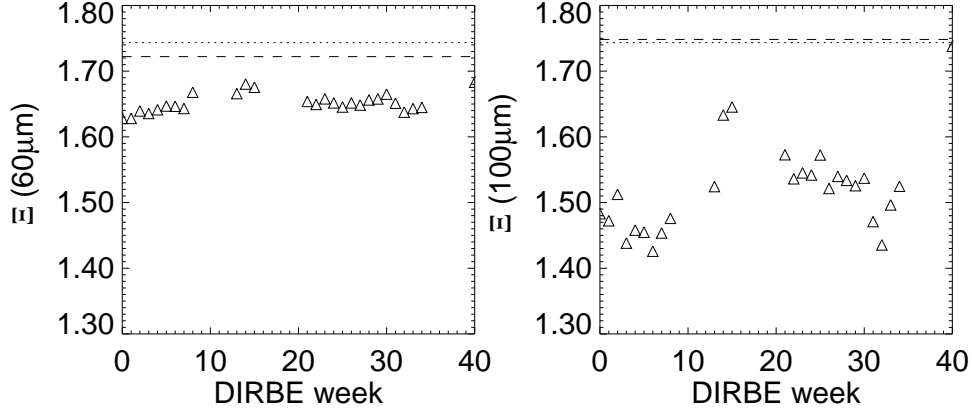}
\caption{Plots of $\Xi(\beta=35\degree)$, as 
defined in eq.~(\ref{equ_xidef}).  In
the absence of a CIB signal, these measurements would agree with
$\Xi_0$ from the Kelsall model (\emph{dashed line}).  A $\csc|\beta|$
(slab) model is overplotted for comparison (\emph{dotted line}).
Plots for other values of $\beta$ are qualitatively similar.  }
\label{fig_zhehplot}
\end{figure}

\BE
\Xi = \frac{B + Z\csc|\beta|}{B+Z} = 
      1+ \frac{\csc|\beta| - Z}{B+Z}
\EE
where $Z$ is $NP+SP$.  Solving for $B$ gives
\BEL{equ_bofxi}
B = \left(1-\frac{\Xi-1}{\csc|\beta| -1} \right) I
\EE
where $I$ is the observed flux at the poles ($Z+B$) and $B$ is the
twice the value of the CIB, as it was in method I.  Figure
\ref{fig_zhehplot} contains plots of $\Xi$ at $35\degree$ for 60 and
$100\micron$  applied to the cirrus-subtracted weekly DIRBE maps.
.  Unsurprisingly, the fit residuals are correlated from
week to week.  To account for this correlation, we estimate there are
no more than 4 independent measurements within the 41 weeks, and thus
the standard deviation of the mean of $\Xi$ is reduced from the rms
scatter by only $\sqrt4$.

Results for method II are shown in Table \ref{table_method2}.  For
each latitude bin and each band $b$, $\Xi_{K_b}$ is calculated from
the Kelsall IPD model and compared with the measured $\Xi_b$.  The
background, $B/2$, is then determined from eq.~(\ref{equ_bofxi}).  A
weighted average gives $28.0\pm 1.9 \nWpMMSr$ at $60\micron$ and
$21.9\pm 2.7 \nWpMMSr$ at $100\micron$.

This agreement between the two methods is encouraging, and suggests
that the observed excess is not coming from within the solar system,
at least it does not vary spatially or temporally in the way the IPD
is expected to.

%------------------------------------------------------------------------------
% Systematic Errors
%------------------------------------------------------------------------------
\section{SYSTEMATIC ERRORS}

\label{sec_errors}
The DIRBE detector noise is small enough that measurement errors in
the determination of the FIR excess are modest, but several distinct
systematic errors contribute to the uncertainty in the final result.
The IPD model dependence is discussed in \S \ref{sec_ipdmodel} and
possible emission correlated with the WIM is considered in
\S\ref{sec_wim}.  Section \ref{sec_xflux}, more generally rules out
emission from a dust slab aligned with the Galactic plane, and
possible emission from the Galactic halo is addressed in
\S\ref{sec_halo}.  Final uncertainty estimates are presented in
\S\ref{sec_err_results}.

\subsection{IPD Model Dependence}
\label{sec_ipdmodel}

A fully self-consistent model of the IPD emission has not yet been found,
probably because of the large number of dust components whose
temperature and density may vary spatially and temporally.  The
Goddard model (\cite{kelsall98}) is certainly the most complete, but
it still must resort to fudge factors to explain the emissivity
function of the IPD, and it assumes that the emissivities of the
spatially separate components are identical.  With currently available
data, it is not economical to introduce still more parameters in order
to solve this problem, so no CIB measurement that depends in detail on
the zodiacal emission model can be trusted.  However, the Kelsall
\etal\ model is used as a reference model in the following. 

Neither method described in the previous section is strongly
influenced by this choice of IPD model.  Our analyses rely only on
data at high ecliptic latitudes, where many of the zodiacal
components, such as the dust bands at $\beta < 15\degree$ can be
safely ignored.  Furthermore, the dimensionless parameters, $R$ and
$\Xi$, can be predicted in a nearly model-independent way and
readily compared with the data.  The advantage of this approach is
that the results depend on relative measurements made on short
timescales, and are almost independent of the choice of IPD model. 
What little dependence there is enters the two methods in different
ways. 

\subsubsection{Method I}

\begin{figure}[t]
\plotone{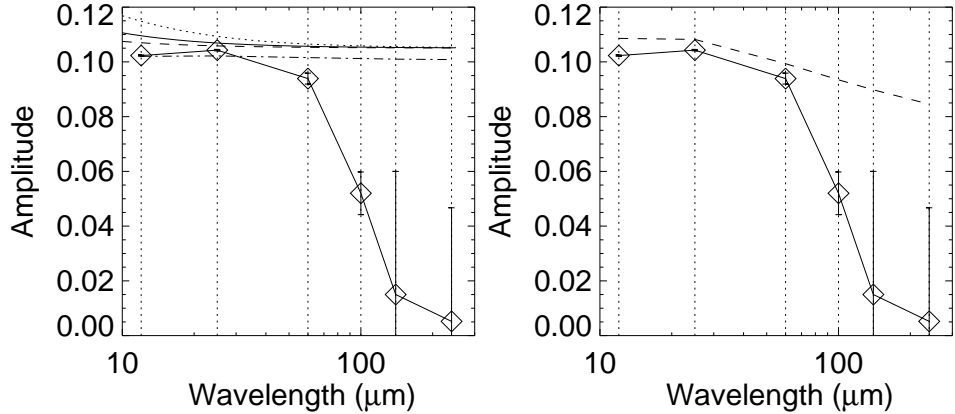}
\caption{Predictions of $A$.
Diamonds are measured amplitudes $A'$ shown in Table
\ref{table_cib_fit} with 1 sigma error bars.  These amplitudes are 
approximately the ratio of emission near the ecliptic plane to the
total emission - and are therefore nearly model-independent.  
a) The solid curve is the theoretical $A$ for the widened fan model
with $T_0 = 286$ K, dashed line is a factor of 2 warmer dust and
dotted line is factor of 2 colder.  
Also shown is the prediction of the Kelsall \etal\ model
(\emph{dot-dashed line}). 
b) Two component model.  Diamonds are measured amplitudes $A'$ as in
previous figure.  Dashed line corresponds to a contrived two-component
dust model as described in the text.  Even in this extreme case, the
model prediction cannot be made to agree with the observation at
$100\micron$.  }
\label{fig_Apred}
\end{figure}

For most reasonable models of zodiacal emission, the expected
amplitude $A$ in method~I should be a weak function of wavelength, not
a constant as we assumed.  Figure \ref{fig_Apred}a, shows the
function $A(\lambda)$ predicted by the Goddard (\cite{kelsall98})
model (\emph{dot-dashed line}) compared to the observed $A'(\lambda)$
values.  The solid line is the Goddard model evaluated at the poles of
the dust plane instead of the ecliptic poles, for $T_0 = 286$K.  In
this model, the dust temperature depends on distance from the Sun, and
$T_0$ is the temperature at 1 AU.  The other lines are for $T_0$
higher and lower by a factor of two.  Of course this is an absurdly
large range of temperatures, but we use it to illustrate that for
single emission components, the amplitude $A(\lambda)$ is nearly
constant on the Rayleigh-Jeans side of the IPD spectrum.  The
dependence of $A$ on other model parameters is similarly weak.

Two-component models of IPD can also be considered.  If there are two
distinct grain sizes with different $T_0$, it is still impossible to
fit the observations.  In the worst-case scenario, the dust grains are
segregated into two populations, one in a dust layer near the ecliptic
plane, and the other away from it.  The spectrum of $A$ is then simply
the ratio of the spectra of the two components - both of which have
identical Rayleigh-Jeans tails.

In order to fit the observed $A'(\lambda)$ values with no background,
one requires two populations of dust grains with different emissivity
laws.  An example of this is shown in Figure \ref{fig_Apred}b - but
this is an extreme case of grain properties and geometry constructed
to give the desired result.  In this model, the IPD away from the
midplane is very similar to the Kelsall model: $T_0=286\K$, blackbody
dust.  However, the slice of dust within $|z|<0.03$AU (the extent of
Earth's annual excursion from the midplane) has $T_0=236\K$ and
$\nu^{0.5}$ emissivity.  These parameters seem unlikely, but they
greatly reduce the excess $60\micron$ flux, and affect the
$100\micron$ excess slightly.

We considered a class of two-component models with different
emissivity power laws, different temperatures, with one in a much
thinner disk than the other.  Exploring such models in detail, we
found that it to be impossible to reduce the $60-100\micron$
background substantially without either producing excessive emission
in the $12\micron$ zodiacal signal or going to unreasonable parts of
the model parameter space.

We therefore proceed by assuming that an extrasolar, isotropic excess
at $60\micron$ is physically more acceptable than a contrived IPD
emission model.  The reasonable range of IPD models do not alter the
derived $60$ and $100\micron$ excesses by more than $5\nWpMMSr$ (95\%
confidence).

\subsubsection{Method II}

Because method II is sensitive to the shape of the $e=90\degree$
emissivity contours (as seen in Fig. \ref{fig_emiscon}), it is immune
to a two-component model of the type contrived above, as long as the
components are layered with azimuthal symmetry.  However, method II is
in general more dependent upon the zodiacal light model used.  This
dependence is calculated by using the Kelsall \etal\ model as a
reference IPD model.  Table \ref{table_kels} displays the error in the
IR excess measurements introduced by $1\sigma$ and $3\sigma$ changes
in the main 6 parameters of the Kelsall model.  Only the parameter
$\alpha$, (the density $\rho \sim R^{-\alpha}$) affects method II
results significantly.  This is not surprising, because $\alpha$
affects the shape of the IPD cloud more than most of the other
parameters.  Uncertainties in the Kelsall model propagate into our
results only at the level of $3\nWpMMSr$ at $60\micron$ and
$1\nWpMMSr$ at $100\micron$ (95\% confidence).  If the Kelsall model
provides a reasonable description of the shape of the IPD cloud, then
the model uncertainties cannot be much bigger than this.  If, however,
the Kelsall model is missing a substantial component that changes the
emissivity contours' shape in the $e=90\degree$ plane, then the
uncertainty could be much larger.  

\subsection{Dust Emission from the ISM}
\label{sec_wim}

In this section we assess the uncertainty in the IR excess described
above due to a zero-point error in the Galactic ISM (or cirrus)
emission maps.  According to SFD98, Table 2, the largest formal
uncertainty in the cirrus emission is less than $1\nWpMMSr$.  However,
there is a larger systematic uncertainty resulting from neglect of
dust in the warm ionized medium (WIM).  The issue is not whether there
is dust in the WIM; all ISM dust emission, described by the SFD98
cirrus maps, is subtracted from the DIRBE weekly maps before $R$ and
$\Xi$ values are fit.  However, any zero point error in the SFD maps
will propagate directly into the measurement of the CIB.  In fact, the
flux measured at the poles contains no cirrus and are essentially a
full-sky zodiacal light fit, evaluated at the poles (method I) and at
other latitudes (method II).  This is important for isotropy
considerations, because the results we obtain effectively use data
from the entire high-latitude sky via the SFD98 fit to the
Leiden-Dwingeloo \HI\ survey.

\subsubsection{Correlation with H-alpha}
One tracer of the WIM that can be used to constrain the cirrus
zero-point is \Halpha\ emission.  A recent paper by
Lagache \etal\ (1999) addresses this question. 
Using the high-quality \Halpha\ data of Reynolds \etal\ (WHAM;
\cite{wham}) to trace the WIM, they claim to find significant WIM
emission at $100-1000\micron$, using 2\% of the sky at high galactic
and ecliptic latitude, and \Halpha\ emission between 0.2 and 2 R.  We
have repeated their analysis with the same data in the same regions of
the sky and find no WIM dust emission uncorrelated with \HI\
emission.  This high-latitude analysis provides no indication that our
zero-point is incorrect. 

Although we are interested in zero point problems in the diffuse
cirrus at high galactic latitude, where the SFD98 zero point was
determined, we must resort to a different analysis that makes use of
\Halpha\ data closer to the galactic plane where the signal is
strong.  This will provide a ``worst case'' result.
Heiles \etal\ have shown that a simultaneous fit of \HI\ and
\Halpha\ yields only a modest shift in zero point (\cite{heiles99}).
In several regions near Eridanus, they perform a fit of the form
\BE
I_\lambda = A + B N({\rm \HI}) + C N({\rm H}\alpha)
\EE
and find that $A$ varies by approximately $6 \nWpMMSr$ RMS at
$100\micron$ in the various regions, with a central value of $2
\nWpMMSr$.  These fits are performed with the SFD98 $100\micron$ map
which already has a model of zodiacal light+CIB removed, i.e., it is
zeroed to \HI.  The fact that these offsets are so near zero indicates
that the dust correlated with \Halpha\ might possibly explain as much as $6
\nWpMMSr$ (the extreme case found by Heiles \etal) of our $\sim30
\nWpMMSr$ background measurement at $100\micron$, but is not likely to
significantly alter the result.

Although the \Halpha\ result is encouraging, it suffers from a few
weaknesses.  The sky coverage is small (2\% in Lagache \etal\,
$\sim10\%$ in Heiles \etal), and future analyses using the entire WHAM
data set may provide more concrete answers. 
Also, the \Halpha\ emission is not proportional to N(\HII) but
rather to 
$\int n_p^2 dl$
and is also weakly dependent on temperature.  This means that a
correlation of \HII\ density and dust/gas ratio could contrive to
produce $A \approx 0$ even though the derived zero point is
incorrect.  Therefore, an alternative method is desirable as
a confidence check.

\subsubsection{Pulsar Dispersion Measures}

The pulsar dispersion measure is a straightforward determination of
the column density of electrons $N(e^{-})$ along the line of sight to
a pulsar.  It does not depend on the temperature or density of the
ionized gas, but does rely on the pulsar being far enough away to give
a fair assessment of the Galactic \HII.  For this work, we have made
use of the pulsar catalogue assembled by J. Taylor at Princeton
(\cite{taylor93}
\footnote{The latest version is available at
{http://pulsar.princeton.edu/ftp/pub/catalog/}.}).

\begin{figure}[t]
\plotone{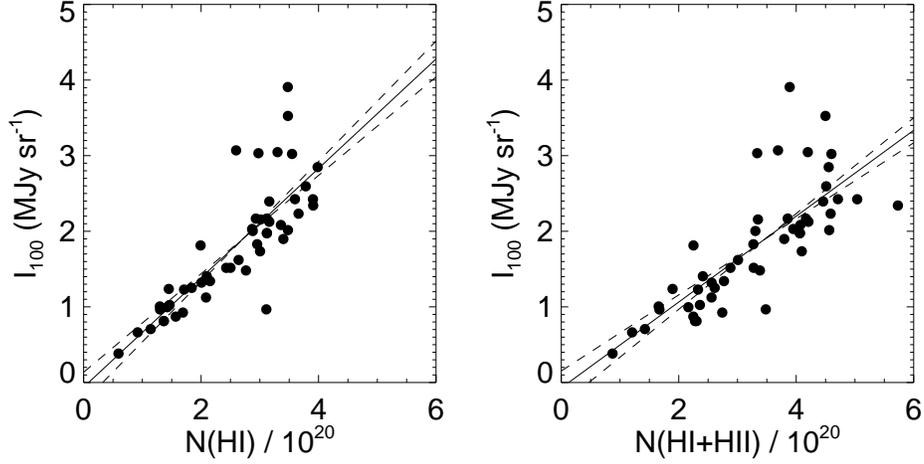}
\caption{The dust/H correlation along the lines of sight to 50
high latitude pulsars ($|b| > 20$, $z > 400\pc$).  Formal $\pm 1\sigma$
errors are shown by the dashed lines. 
a) Dust vs. $N$(\HI). 
b) Dust vs. $N$(H).  $N$(\HII) derived from pulsar DMs is added to N(\HI). }
\label{fig_pulsar}
\end{figure}

The catalogue contains 707 pulsars, 146 of which are at $|b| >
20\degree$ and have distance quality codes of ``a'' or ``b''.  Of
these, 108 are in the region covered by the Leiden-Dwingeloo survey,
allowing a comparison of $N(e^{-})$ and \HI\ with $100\micron$ flux.
A further requirement that the pulsars be out of the plane ($|z| >
400$pc) reduces the list to 46.  As one can see in Figure
\ref{fig_pulsar}, the zero point of the H/dust regression changes only
modestly when the pulsar data are included, even though the slope
changes by roughly 1/3.  This indicates that \HI\ and \HII\ are
correlated, but perhaps are no more correlated than any other $\sim
\csc|b|$ mechanism.  Because the scatter is no tighter with \HII\
included, one might conclude that there is little dust associated with
\HII.  On the other hand, the poor assumption that each pulsar is
behind all Galactic dust may add noise, cancelling out the
improvement. 

Curiously, the $y$-intercept in \ref{fig_pulsar}(a) is not zero, even
though it is forced to be (by construction) over the average
high-latitude sky (see SFD98).  This nonzero intercept may indicate
variation in the gas/dust ratio - even at high $|b|$, or may also
reflect large-scale gradients in the SFD98 dust model.  Whether the
SFD98 temperature correction is used or not, we find that the largest
zero-offset induced by such changes is at most $0.1\MJypSr$ ($3
\nWpMMSr$) at $100\micron$.  If we assume that perhaps half of the
dust emission emanates from above the 400pc pulsar cut, and double
this effect, it is still negligible.  Furthermore, if we consider the
change in zero point due to the use of a temperature correction as a
systematic error that propagates directly to the IR excess, we still find a
systematic uncertainty of only $6 \nWpMMSr$.  Because this is an
uncertainty similar to that obtained from the
\Halpha\ analysis of Heiles above, we adopt $6 \nWpMMSr$ as the (95\%
conf.)
uncertainty associated WIM-correlated dust emission, and add this to
our systematic error budget (Table \ref{table_errors}) at 60 and
$100\micron$.

\subsection{Ruling out a dust slab}
\label{sec_xflux}

There is still a chance that a diffuse layer of dust more than 400pc
above the disk of the Galaxy could be responsible for the emission.
Such a layer, if behind most of the pulsars, could be either
uncorrelated with \Halpha\ emission, or could be associated with \HII\
so diffuse that \Halpha\ emission is effectively suppressed.  This
sort of a foreground would be indistinguishable from the IR excess using the
methods described in this paper, but would reveal itself by a
dependence on Galactic latitude.

Because the DIRBE data do not extend over a full year, and because of
drifts in the zodiacal light intensity and detector gain with time,
the annual average maps contain unphysical gradients that may confuse
a direct fit of a $\csc|b|$ component.  In order to test for the
presence of such a component, we again introduce a dimensionless
parameter.

For each weekly DIRBE map, we construct a dimensionless parameter
$\chi$ from the flux in four patches on the sky, always placed at
solar elongation $|e|=90\degree$ and $|\beta| = 75\degree$.

The mean flux values in the four patches of sky are designated
$I_{NF}$ for ``North-Forward'', $I_{NB}$ for ``North-Backward'', and
likewise $I_{SF}$, and $I_{SB}$ for the south, just as in the
definition of $\Xi$.  In this case, ``North forward'' refers to a
direction on the sky $(\lambda, \beta) = (L_{true}+90, +75)$ where
$L_{true}$ is the true heliocentric longitude of Earth.  The
$L_{true}+90$ direction does not correspond precisely to the direction
of Earth's velocity around the Sun because of eccentricity; rather it
is at solar elongation $90\degree$.  These values are computed for
each week of data, except when any of the patches is at low Galactic
latitude ($|b| < 10\degree$)

The four lines of sight used each week form an ``X'' in space.  Two
useful combinations are $I_{A} = I_{NF}+I_{SB}$ and $I_B =
I_{NB}+I_{SF}$.  We then define the dimensionless ratio
\BEL{equ_chi}
\chi = \frac{I_A-I_B}{I_A+I_B}
\EE
in which gain drifts and dust variations nearly cancel out.  Because
$I_A$ and $I_B$ are measured at solar elongation $90\degree$, $\chi$
would be zero in the absence of ISM emission, if the IPD were aligned
with the ecliptic plane.  Misalignment with the ecliptic plane will
produce a periodic signal in $\chi$, revealing the inclination of the
dust plane if the ascending node is known.  Likewise, a uniform slab
of emitting material will contribute another $\csc|b|$ periodic term.
Unfortunately, the Galactic plane and dust midplane have similar ascending
nodes so that their signal in this statistic is nearly degenerate,
making a simultaneous fit impossible.  Fortunately, an error of
$1\degree$ in the inclination of the dust plane would result in an error
of 1.5 and $0.3\nWpMMSr$ at the poles in the 60 and $100\micron$
channels respectively.  An isotropic background does not
contribute to $\chi$.

\begin{figure}[t]
\plotone{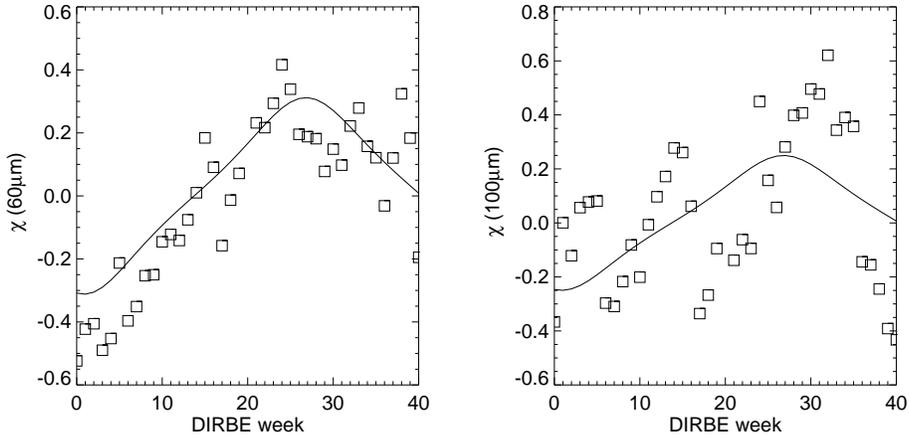}
\caption{Test for $\csc|b|$ slab of dust.  (a) Observed value of
$\chi$ at $60\micron$ as defined in eq. (\ref{equ_chi}).  The solid
line is the $\chi$ predicted from a $\csc|b|$ model.  (b) $\chi$ at
$100\micron$.  }
\label{fig_xflux}
\end{figure}

As an example of the power of this technique, we display 
in Figure \ref{fig_xflux} $\chi_{60}$
and $\chi_{100}$ for the 60 and $100\micron$ channels after removal of
the \HI\ correlated component.  For these fits,
we used the inclination angle $i=2.03\degree$ obtained by the DIRBE
team (\cite{kelsall98}).  Before \HI\ removal (not shown in figure)
the $\csc|b|$ term is strong, with values of 40 and $100 \nWpMMSr$ at
the Galactic poles for 60 and $100\micron$ respectively.  The flux is
twice that at the ecliptic poles at $|b|=29.8\degree$.  After removal
of the best fit \HI\ coefficient, only a small signal is left, at the
level of 2.5 and $1.2\nWpMMSr$.  Even though we have not explicitly
removed a $\csc|b|$ component, the \HI\ fit appears to have done so.
The same procedure is difficult in the other wavebands.  The noisy 140
and $240\micron$ channels do not give a meaningful measurement of
$\chi$, and at 12 and $25\micron$ the technique is much more sensitive
to the inclination angle of the dust plane.

This small residual signal at 60 and $100\micron$ may reflect a small
error in the \HI\ subtraction, or may actually be an emission
component not correlated with \HI.  Whether real or not, this
component is very small compared to the IR excess measured in this
paper, and in fact is small compared to the WIM error derived above.
Therefore, we consider the error due to a Galactic slab to be already
included in the WIM error adopted above. 

\subsection{Galactic Halo Dust}
\label{sec_halo}

In this section we consider another potential component of Galactic
emission which would not have appeared in this other tests: dust
emission in the halo of our galaxy.  We have ruled out dust correlated
with \Halpha, dust correlated with \HII\ within 400pc of the Galactic
plane, and dust in a roughly $\csc|b|$ distribution.  However, a very
diffuse component of dust mixed into the Galactic halo -- associated
with ionized H or perhaps no gas at all -- has not been strictly ruled
out.  The marginal detection of reddening along lines of sight passing
near spiral galaxies (\cite{zaritsky94}) implies that there may be
dust at $r>60$kpc, although it is another question whether such dust
is warm enough to mimic the measured excess.  Alton \etal\ (1998) have
observed cold
(15K) dust 2 kpc off the disk of NGC 891 with SCUBA, but such dust
would not give the observed signal at 60 and $100\micron$ (see also
Howk 1999).

Mechanisms that transport dust from the Galactic disk into the halo by
radiation pressure have been proposed (e.g. \cite{ferrara91}), but
contain only crude approximations to the Galactic magnetic field.
Consideration of a more realistic magnetic field would increase the
diffusion time and increase the probability of dust destruction.  The
dust which makes it into the halo, if any, might be expected to be
hotter than the disk dust, as long as the radiation field is similar.
The grains should be smaller, and thus attain a higher equilibrium
temperature; and the radiation pressure mechanism preferentially
transports grains with a high optical/UV cross section to mass ratio -
and these may also tend to be hotter.  However, to explain the
observed spectrum, they must be quite hot.  Taking the warm
$\nu^{2.6}$ component from Finkbeiner \etal\ (1999), a temperature of
$\sim 28\K$ must be maintained to explain the observed spectrum - in
contrast to an average interstellar dust temperature of $18\K$ for
single-component $\nu^2$ models or $16\K$ for the warm component of
the Finkbeiner \etal\ two-component model.

Such a halo must also be nearly isotropic.  The halo proposed by
Ferrara \etal\ (Fig. 2 in \cite{ferrara91}) is unlikely, as it would
have been detected by the method described in \S\ref{sec_xflux}.  In
fact, a uniform spherical dust halo bright enough to explain the IR
excess would appear isotropic enough for $r_{halo} \ga 20$kpc that it
would not be noticed in any conceivable analysis of the DIRBE data.
The theoretical complications of such a halo are many, but are perhaps
no more distasteful than the conflict this IR background causes with
TeV gamma ray observation, as we discuss in \S\ref{sec_tev}.  
However, the idea appears to be ruled out by the 
observational non-detection of a $60\micron$ halo in M31 at the level
of $\sim 0.1\MJypSr$ ($5\nWpMMSr$).  In fact, Rice \etal\ (1988) found
no evidence that the IRAS $60\micron$ emission of optically large
galaxies extends into the halos.  It is difficult to see why our
Galaxy should be any different. 

To summarize this section, if there is indeed a component of dust in
our Galaxy that accounts for the IR excess, it must be uncorrelated
with \Halpha, more than 400pc off the plane, and differ substantially
from $\csc|b|$.  This forces us to seriously consider the possibility
that the observed excess is mostly extragalactic in origin.

\subsection{Final Results}
\label{sec_err_results}

In this section, we average the results from methods I and II and
state the final results for the DIRBE 60 and $100\micron$ channels,
including systematic uncertainties.  The casual reader should skip directly
to \S \ref{sec_combined}

\subsubsection{Method I}
In method I, several computational choices introduce systematic
uncertainties.  One choice is whether to take a mean, median, or
mid-average of the pixels at the ecliptic poles in each week.  Some
outlier rejection is necessary because of imperfect point source
removal.  However, because there is a gradient across the polar cap
(due to solar elongation dependence) a mean is more stable.  All
results in this paper were derived with a mid-average, in which the
highest and lowest 10\% of the values are discarded, and the remaining
values averaged.  This proves to be more robust than either a mean or
a median.  Use of a mean increases our results by $\sim 5\nWpMMSr$ at
$60\micron$ and twice that at $100\micron$.  However, a straight mean
also results in an unacceptable $N-S$ asymmetry (formally $25\sigma$
at $25\micron$) which is not present when a mid-average is used.

Another choice is the size of the polar region.  Ecliptic latitude
cuts of $|\beta|=85\degree$ and $|\beta|=87\degree$ were tested.  The
$85\degree$ cut provides adequate signal, but excludes the brightest
parts of the LMC.  The variation introduced by this choice is
negligible at 60 and $100\micron$, but large at 140 and $240\micron$.
This is because the uncertainty in those channels is dominated by
measurement noise, and a $|\beta|=87\degree$ leaves much less signal
to work with. 

The use of the ecliptic pole is motivated by the obvious symmetries.
Another potential choice is the apparent pole of the dust plane, which
is inclined $\sim 2\degree$ with respect to the ecliptic.  This choice
can modify results by $5\nWpMMSr$.

A number of pixels are discarded for lack of coverage.  Some weeks of
DIRBE data, particularly ``science'' weeks 20 and 35 (numbered 0 to
40) contain very little sky coverage and are rejected.  Other weeks
with more than 10\% bad pixels at the poles are also discarded
so as not to mask out too many pixels in the remaining weeks.

The masks are generated independently for each waveband, which may
raise questions about comparing the derived $A'$ and $C$ parameters
for each waveband.  However, the amplitude $A$ determined at $12,
25\micron$ is quite insensitive to the mask used.  One can take the
union of multiple masks and apply that ``master'' mask to each
waveband and still obtain the same results, though with lower
signal-to-noise.  We have no indication that such a procedure is
necessary, so we simply apply to each waveband its own mask.  Table
\ref{table_pixmask} shows the fraction of pixels lost due to the mask
in each waveband, and the number of weeks of good data used.

Even the method of pixel mask symmetrization calls for some
judgment.  For each bad pixel $(\lambda, \beta)$, it is necessary to
mask $(\lambda, -\beta)$, or else the solar elongation gradient will
be aliased into the time domain, and appear as signal in $R$.  In
fact, masking $(-\lambda, -\beta)$ and $(-\lambda, \beta)$ corrects
this problem to a higher order.  This 4-fold mask symmetrization was
followed for all Method I results.  Failure to symmetrize the mask in
this way adds $2\nWpMMSr$ at $60\micron$ and $6\nWpMMSr$ at
$100\micron$.

In light of these systematic errors, we assign a systematic
uncertainty of 7 (11)$\nWpMMSr$ at 60 (100)$\micron$ to method I (95\%
conf.), in addition to the formal statistical errors.  At 140 and
$240\micron$, measurement noise dominates, so no such systematic
errors are added.

The intent of method I is to make reliable IR excess measurements at
$60-240\micron$, but the less reliable results obtained at shorter
wavelengths are interesting as well.

The flux measured at $12\micron$ ($\sim2\%$ of zodiacal light) is
almost certainly an artifact of the greater model dependence at short
wavelengths.  The $5\micron$ emission shown in Table
\ref{table_cib_fit} is relatively stronger
($\sim10\%$ of zodiacal light) but is subject to extreme model
dependence.  It is interesting that this flux level of $\sim
23\nWpMMSr$ is similar to the $5\micron$ excess found by Dwek \&
Arendt (1998).  They declined to call this a CIB because of
anisotropy.  This agreement between the two numbers may only be a
curious coincidence. 

\subsubsection{Method II}
In method II, there are fewer choices to make.  Each line of sight
uses pixels in a $5\degree$ diameter patch, as this is the largest
patch that will be statistically independent from week to week and
latitude bin to latitude bin.  This method is sufficiently robust that
the choice of the ecliptic as the symmetry plane is unproblematic: the
Kelsall model gives less than 1 part in 1000 variation in $\Xi$ due to
this choice.  In fact, the only significant systematic error this
method has in common with method I is the mask-symmetry error, which
we take to be the same as in method I.  We therefore adopt a
systematic uncertainty of 2 (6)$\nWpMMSr$ at 60 (100)$\micron$ for
method II.

\subsubsection{Combined Results}
\label{sec_combined}
The two methods are complementary: the first uses time-variability at
the ecliptic poles for the analysis, and the second uses the spatial
morphology of the data in each week.  Method II is superior in that it
samples a larger fraction of the data set and achieves a much higher
S/N, but the use of different regions of the sky in each week results
in systematic errors that are difficult to understand in detail.
Therefore, there is no \emph{a priori} reason to prefer one method
over the other.  We combine the results with a weighted average, using
the formal random errors and systematic errors that apply to each
method.  Such errors average down when results of the two methods are
combined (see Table \ref{table_errors}.  The remaining systematic
errors that apply to \emph{both} methods, such as the uncertainty in
the ISM emission, are added after this averaging.

In Table \ref{table_results} the results for method I, method II, and
the average are stated.  These numbers represent the FIR excess in 60
and $100\micron$ DIRBE filters, which are calibrated assuming a flat
spectrum in $\nu I_\nu$.  Color corrections for plausible FIR
background spectra do not change these results by more than $10-20$\%
(see \cite{dirbesupp95}).

%------------------------------------------------------------------------------
% Discussion
%------------------------------------------------------------------------------
\section{DISCUSSION}
\label{sec_discussion}
Because we are unable to find any emission component within the Galaxy
to explain the observed IR excess, we tentatively interpret the
emission as an isotropic extragalactic background (CIB).  In this
section we discuss the consequences of such an interpretation.

\subsection{The EBL Energy Crisis}
\label{sec_crisis}

The integrated energy in the CIB for $60\micron < \lambda < 1$mm is
$40\pm12\nWpMMSr$ according to our measurements and those of Fixsen
\etal\ (1998) - roughly twice the
FIR energy content inferred from Hauser \etal\ (1998), and much higher
than that predicted by Malkan \& Stecker (1998).  Where is this
energy coming from? 
A comprehensive analysis by Bond, Carr, \& Hogan (1986) reviews
several possible sources for the FIR background, including primeval
galaxies, pregalactic stars, black hole accretion, and decaying
particles.  Nuclear fusion within stars at epoch $z$
contributes a radiation energy density $\Omega_R$ today (in units
of the critical density) of 
\BE
\Omega_{R}\approx 0.007 {\Delta Z \over 1+z}\Omega_B {\cal F}
\EE
where $\Omega_B$ is the  baryon density of the Universe today
(e.g. $\Omega_B h^2 = 0.019 \pm 0.001$; \cite{burles98}),  $\Delta Z$ is
the mean metallicity of {\it all} baryons in the Universe, and ${\cal F}$
is the fraction of the emitted radiation which is reprocessed into the
FIR.

Black hole accretion is another source of energy for the CIB and yields
a radiation density  
\BE
\Omega_{R}= {\left(\epsilon\over 1-\epsilon\right)} (1+z)^{-1} \Omega_{BH,acc}
{\cal F} 
\EE
where $\epsilon\approx 0.1$ is the efficiency of rest mass to energy
conversion 
and $\Omega_{BH,acc}$ is the mass density of accreting black holes. 

With the definition of the critical density,
$\rho_{crit}=1.9\times10^{-29}h^2 {\rm g}\cm^{-3}$ and $h=H_0/$(100 km
s$^{-1}$ Mpc$^{-1})$, the detected background of $40\nWpMMSr$ becomes
$\Omega_{CIB}=9.8\times10^{-7}h^{-2}$. If this radiation was generated
within stars and mostly reprocessed into the FIR at $z\approx 2$, then
$\Delta Z \approx .02$, (regardless of $h$) averaged over the entire
Universe.  This seems rather large, and would indicate that most star
formation must be well hidden from view.

Another possibility  is generation of the CIB by accretion onto black holes, 
again
at $z\approx 2$, with most of the radiation emerging in the FIR.
  If the black holes were formed by
the accretion of baryonic matter ($\Omega\approx 0.019h^{-2}$), with
an efficiency $\epsilon=0.1$, then
the cosmological density of black holes today must be $\Omega_{BH}
\approx 2.9\times10^{-5}h^{-2}$
and the fraction of all baryons that
must have fallen into massive black holes (assuming no accretion of dark 
matter) is approximately 0.15\%.  While this number is 
large, it is smaller than the black hole mass fraction of 0.5\% in
galactic bulges (\cite{magorrian98}).

It is also possible that energy extraction from black hole accretion
is more efficient than assumed above.  Recent work by Gammie (1999)
indicates that for rotating black holes with a magnetic field,
accretion efficiencies of $\epsilon \approx 0.5$ are possible.  This
reduces the requirement to 0.015\% of baryon mass in black holes,
consistent  with the Magorrian
\etal\ measurement and a disk/bulge mass ratio of $\sim 30$. 

Recent work (\cite{almaini99}) shows that AGN observed by Beppo-Sax
can explain the 30 keV X-ray background, and concludes that most of
the energy generation takes place in obscured AGN.  However, those AGN
have only been shown to contribute 10-20\% of the extragalactic
background at 240 and $850\micron$, and are unlikely to produce the
measured flux at 60-$100\micron$.  A recent model by Fabian (1999)
suggests that the majority of black holes undergo a highly obscured
growth phase, and predicts a population of objects at $z>1$ which emit
predominantly hard ($E>30\keV$) X-rays and FIR/submillimeter photons.
The Chandra telescope has recently resolved the majority of the 2-10
keV background, finding two new classes of objects: 1) optically
``faint'' galaxies ($I >> 23$) with very high X-ray to optical ratios,
and 2) point-like, luminous hard X-ray sources in the nuclei of normal
bright galaxies showing no other sign of activity
(\cite{mushotzky2000}).  The former group is consistent with either
early quasars, or extremely dust enshrouded AGN at $z>2$.  Conceivably
some combination of accretion onto black holes and powerful starbursts
within the core regions of merging galaxies is capable of explaining
the IR excess reported here.

The spectral shape of the IR excess also has interesting
implications.  Because the $60/100\micron$ ratio is $\sim 1$ (in $\nu
I_\nu$ units) the source must have a high dust temperature of $T
\approx (1+z) 28\K$.  For $z\approx 2-3$ this temperature indicates a
violent process is at work, and argues in favor of a few hot, bright
sources.  SIRTF will provide valuable information about the
luminosity function of such sources at 24, 70, and $160\micron$.

\subsection{TeV Gamma Crisis}
\label{sec_tev}

Another observational conflict is inevitable if the observed IR excess
is of extragalactic origin: the observation of gamma rays 
by HEGRA (\cite{konopelko99}, \cite{aharonian99}) at energies
up to $E \approx 20$TeV from Mkn 501, and by 
Whipple (\cite{samuelson98}) up to $E \approx 10$TeV.
The $\gamma$-ray opacity on CIB photons may be approximated by 
\BE
\tau_{\gamma\gamma}(E_{\gamma}) \approx
0.24\left(\frac{E_\gamma}{1\TeV}\right)
\left(\frac{u(\epsilon_*)}{10^{-3}\eV\cm^{-3}}\right)
(z_S/0.1)h^{-1}_{60}
\EE
where $u(\epsilon_*)=\epsilon_*^2n(\epsilon_*)$ is the typical energy
density in an energy band centered on $\epsilon$, $h_{60}$ is the
Hubble constant, and $z_s$ is the source redshift (see
\cite{coppi99}).  Integrating the exact cross section over the DEBRA
(Diffuse Extragalactic Background Radiation, including CMBR, CIB and
optical/UV EBL), Stecker \& de Jager (1998) find an optical depth of
2.5 at $20\TeV$ photons using a CIB prediction (Malkan \& Stecker
1998) based on an extrapolation of \IRAS\ counts and other data.  The
HEGRA data are consistent with this value at 20 TeV
(\cite{konopelko99}).  However, using the currently accepted
measurements of the CIB, Coppi \& Aharonian (1999) obtain an optical
depth $\tau\approx5$ for $20\TeV$ photons.  If we modify the CIB
spectrum to take account of the 60-$100\micron$ CIB measurements given
in this paper, $\tau$ doubles to $\sim10$.  Even without this doubling, the
TeV observations imply that the intrinsic Mkn 501 spectrum is concave
upward in the 10-30 TeV range, contradicting the synchrotron
self-Compton emission model which requires it to be concave downward.
Including the current CIB measurements would imply the intrinsic Mkn
501 spectrum increases a factor of 1000 in $\nu F_\nu$ from 10 to
$20\TeV$.  Although very little is yet known about the true intrinsic
spectrum of these blazars, this seems an unlikely explanation
(however, see Mannheim 1999 for an alternative emission model).

There has also been speculation (\cite{harwit99}) that multiple TeV
photons may be emitted coherently by blazars such as Mkn 501, and
might arrive in Earth's atmosphere so close in time and space that
they are confused with a single-photon event.  Such coherent emission
seems implausible, given the very large phase space available to TeV
photons, but some similar mechanism might yet resolve the apparent conflict
between TeV gamma observations and the expected opacity of the CIB. 

Still more radical explanations have been proposed.  Coleman \&
Glashow (1997, 1999) have proposed that quantum gravity effects cause
a small violation of the invariance principle for very high energy
particles.  This effect may be large at the Planck scale
($10^{19}\GeV$) but even at 20 TeV could have measurable consequences.
Kifune (1999) has shown that one possible effect of such a violation
is a sudden drop in the effective $\gamma$-ray / IR cross section for $E
> 10\TeV$.  The resulting dispersion relation may also have observable
consequences for $\gamma$-ray bursts at cosmological distances
(\cite{amcam98}).  Future TeV $\gamma$-ray data may rule out these
quantum gravity theories, or perhaps demand a further exploration.

\subsection{Future Data}

Data to be gathered within the next few years should be adequate to
resolve the problems discussed in this paper.  The Space Infra-Red
Telescope Facility (SIRTF) will obtain deep number counts at 25, 70,
and $160\micron$ with the Multiband Imaging Photometer (MIPS)
(\cite{rieke96}).  If the integrated flux from measured sources sums
to the observed IR excess, then the TeV $\gamma$-ray measurements must
be reinterpreted.  However, if the source density of objects
generating the CIB is sufficiently high, then the deep SIRTF images
will be confusion limited, and constraints on this diffuse emission
will have to come from a fluctuation analysis, in which separation of
the extragalactic component from the diffuse cirrus emission of our
galaxy will be a limiting factor.  Non-detection by SIRTF would
confine the emission to our Galaxy, indicating either Galactic halo
dust emission, a serious systematic calibration problem in the
COBE/DIRBE instrument, a new emission component, or a serious flaw in
the present analysis.

Another test is to observe more blazars at still higher
energies.  By combining data from blazars at different distances and
different energies, one can place reliable limits on the DEBRA
intensity, as long as the intrinsic blazar spectra are at all
similar, or can be predicted by their relation to the X-ray emission
from these objects.  Work by the HEGRA and Whipple
collaborations is underway.

%------------------------------------------------------------------------------
% CONCLUSIONS
%------------------------------------------------------------------------------
\section{SUMMARY}
\label{sec_summary}

Previous attempts by the DIRBE team to measure the cosmic IR
background made use of a sophisticated model of the Inter-Planetary
Dust (IPD) (\cite{kelsall98}).  Such detailed modeling was necessary
to establish the isotropy of the CIB signal detected at 140 and
$240\micron$.  The DIRBE team found excess emission at $100\micron$
also, but doubted the isotropy of the emission and declined to call it
a measurement.  In order to recover the valuable information about
galaxy formation and evolution contained in the CIB, we have
measured the FIR excess at 60 and $100\micron$. 

We analyze the excess DIRBE emission using two different methods.
Each of these methods uses a dimensionless parameter derived from the
DIRBE data in each week of the mission, parameters that are robust
with respect to dust temperature and density variation and detector
gain drift.  These statistics are nearly insensitive to details of the
IPD model.  It is not necessary to know the IPD emission for every
line of sight at every time, although we use the Kelsall \etal\ (1998)
model as a reference.  Method I uses time variation observed in the
flux at the ecliptic poles to measure the background at the poles.
Method II uses the spatial ``shape'' of the $e=90\degree$ data for
each week to remove it and yields an independent measurement of the
background in each week.  Results derived from these two methods are
consistent with each other, giving a background of $\nu I_\nu = 28.1
\pm 1.8 \pm 7\nWpMMSr$ at $60\micron$ and $ 24.6 \pm 2.5 \pm
8\nWpMMSr$ at $100\micron$.

A variety of arguments rule out alternative sources of emission.
Analyses of pulsar dispersion measures and WHAM \Halpha\ data
demonstrate that signal from the WIM-correlated dust is already
accounted for in the cirrus zero point determined from \HI.  After
removing the cirrus, there is no additional component of this emission
correlated with $\csc|b|$, arguing against any additional dust slab
aligned with the Galaxy.  The absence of extended far-IR emission
halos around nearby galaxies (e.g. M31) rules out dust emission from
extended halos.  Unable to find an alternative emission mechanism, we
cautiously consider the implications if this excess is an
extragalactic background.

The energy required to produce a $44 \pm 9 \nWpMMSr$ integrated FIR
background is large compared to the energy expected from stellar
fusion.  If the observed flux is indeed of extragalactic origin, then
stellar fusion is probably not the dominant source of energy in the
universe.  Alternative sources such as highly obscured AGN at moderate
redshift are a possibility, but would predict that 0.15\% of all
baryons are in black holes at the present time (although this figure
is uncertain by a factor of $\sim 10$ because of uncertainty in
accretion efficiency).  A large X-ray background would also be
predicted, unless the obscuration is of sufficient optical depth
($N(H) > 10^{24}$) to block it.  The most serious problem with an
extragalactic origin of the IR excess is the observation of TeV gamma
rays.  The opacity of the measured CIB to 20~TeV photons coming from
Mkn 501 is 10 optical depths, much greater than the apparent
absorption measured by HEGRA (see \S \ref{sec_tev}).  Because of these
inconsistencies, there is currently no satisfactory explanation for
the observed excess, especially at $60\micron$.  We continue to search
for possible sources of emission in the solar system or Galaxy that
could account for the observed emission, and urge caution in the use
of these results.  We eagerly await source counts from SIRTF and the
X-ray observatories that might help to solve this mystery.

%------------------------------------------------------------------------------
% ACKNOWLEDGMENTS
%------------------------------------------------------------------------------
\section{ACKNOWLEDGMENTS}
We would like to thank Carl Heiles, Chris McKee, Bill Reach, Rick
Arendt, and Eli Dwek for helpful discussions.  Computers were
partially provided by a Sun AEGP Grant.  DJS is partially supported by
the \MAP\ project and by the Sloan Digital Sky Survey.  This work was
supported in part by NASA grants NAG 5-1360 and NAG 5-7833.  The
\COBE\ datasets were developed by the NASA Goddard Space Flight Center
under the guidance of the \COBE\ Science Working Group and were
provided by the NSSDC.

%------------------------------------------------------------------------------
% REFERENCES
%------------------------------------------------------------------------------

\bibliographystyle{unsrt}
\bibliography{gsrp}

%------------------------------------------------------------------------------
% FIGURES
%------------------------------------------------------------------------------

%----------------------------------------------------------------------

%----------------------------------------------------------------------

%----------------------------------------------------------------------
% 
% \begin{figure}[t]
% \plotone{zodcompare.ps}
% \caption{Comparison of Goddard and SFD zodiacal light models.
% The difference of the Goddard zodiacal light model
% (\cite{kelsall98}) and the SFD98
% zodiacal light model at $60\micron$ (\emph{a}), $100\micron$ (\emph{b}),
% $140\micron$ (\emph{c}), and $240\micron$ (\emph{d}).  
% The error bars are the RMS dispersions of values in each latitude bin.
% The sense of the difference is that the SFD98 zodiacal-subtracted
% $100\micron$ map is lower by $\sim 1\MJypSr$. 
% }
% \label{fig_zodcompare}
% \end{figure}
%----------------------------------------------------------------------

%----------------------------------------------------------------------

%----------------------------------------------------------------------

%----------------------------------------------------------------------

%----------------------------------------------------------------------

%----------------------------------------------------------------------

%----------------------------------------------------------------------

%----------------------------------------------------------------------

%----------------------------------------------------------------------

%----------------------------------------------------------------------

%----------------------------------------------------------------------

%----------------------------------------------------------------------

%----------------------------------------------------------------------

%----------------------------------------------------------------------

%----------------------------------------------------------------------

%----------------------------------------------------------------------

%----------------------------------------------------------------------

%------------------------------------------------------------------------------
% TABLES
%------------------------------------------------------------------------------
% Table 1

\clearpage
\begin{deluxetable}{crrrrrrr}
\tablewidth{0pt}
\tablecaption{CIB Results: Method I
   \label{table_cib_fit}
}
\tablehead{
   \colhead{$\lambda$}    &
   \colhead{$A'$}    &
   \colhead{$\nu_0$} &   
   \colhead{$C$}  &  
   \colhead{$I$}  &  
   \colhead{$B_N$}  &  
   \colhead{$B_S$}    \\
   \colhead{($\micron$)}  &
   \colhead{($\times 10^{-4}$)}  &
   \colhead{(deg)}  &
   \colhead{($\times 10^{-3}$)}  &
   \colhead{(nWm$^{-2}$sr$^{-1}$)}  &
   \colhead{(nWm$^{-2}$sr$^{-1}$)}  &
   \colhead{(nWm$^{-2}$sr$^{-1}$)}
}
\startdata
  5 &$   947.\pm  14.6 $&$ -14.9\pm   0.8 $&$  -6.9\pm   1.0 $&$  492.8 $&$  21.0\pm   3.7 $&$  24.4\pm   3.7 $\\
 12 &$  1023.\pm   2.5 $&$ -26.9\pm   0.1 $&$  -1.6\pm   0.2 $&$ 6214.2 $&$  55.5\pm   7.9 $&$  65.2\pm   7.9 $\\
 25 &$  1043.\pm   2.6 $&$ -25.6\pm   0.1 $&$  -0.6\pm   0.2 $&$ 4930.9 $&$  -1.0\pm   6.7 $&$   2.0\pm   6.7 $\\
 60 &$   939.\pm  19.8 $&$      ...       $&$  -3.3\pm   1.4 $&$  583.2 $&$  28.1\pm   5.9 $&$  30.0\pm   5.9 $\\
100 &$   520.\pm  78.3 $&$      ...       $&$   1.0\pm   5.8 $&$  161.4 $&$  40.5\pm   6.5 $&$  40.4\pm   6.5 $\\
140 &$   150.\pm 451.5 $&$      ...       $&$   7.5\pm  31.4 $&$   85.5 $&$  36.9\pm  19.9 $&$  36.3\pm  19.9 $\\
240 &$    52.\pm 414.5 $&$      ...       $&$   7.8\pm  28.8 $&$   30.8 $&$  14.8\pm   6.6 $&$  14.5\pm   6.6 $\\
\tablecomments{Fit parameters described in the text.  
$A'$ is the amplitude of the annual variation in the dimensionless
ratio $R$.  The ascending
node true anomaly, $\nu_0$, is fit for 12 and $25\micron$, and
forced to be $\nu_0=-25.0\degree$ for the other wavebands for fit
stability.  
$C$ is the mean value of $R$ averaged over an entire cycle.  $I$ is
the total emission at the poles $(B+Z)$. 
Values of $B_N$ and $B_S$ (excess at north and south
ecliptic poles) are shown separately to demonstrate that the model is
robust with respect to a $N-S$ difference in the cirrus. 
}
\enddata
\end{deluxetable}

%----------------------------------------------------------------------
% Table 2

\clearpage
\begin{deluxetable}{r|rrrr|rrr}
\tablewidth{0pt}
\tablecaption{CIB Results: Method II
   \label{table_method2}
}
\tablehead{
   \colhead{$\beta$}      &
   \colhead{$\csc\beta$}  &
   \colhead{$\Xi_{K_{60}}$}      &
   \colhead{$\Xi_{60}$}        &
   \colhead{$B_{60}$/2}     &
   \colhead{$\Xi_{K_{100}}$}      &
   \colhead{$\Xi_{100}$}      &
   \colhead{$B_{100}$/2}  \\
   \colhead{(deg)}  &
   \colhead{}  &
   \colhead{}  &
   \colhead{}  &
   \colhead{($\nWpMMSr$)}  &
   \colhead{}  &
   \colhead{}  &
   \colhead{($\nWpMMSr$)}  
}
\startdata
%Results for      60.0000 microns                         100 microns
35 & 1.743 & 1.722 & 1.649 & $ 29.3 \pm  2.9 $ &  1.748 & 1.524 & $ 23.4 \pm  3.7 $\\
40 & 1.556 & 1.553 & 1.501 & $ 27.2 \pm  3.3 $ &  1.571 & 1.416 & $ 21.2 \pm  5.3 $\\
45 & 1.414 & 1.419 & 1.378 & $ 28.1 \pm  4.9 $ &  1.432 & 1.312 & $ 21.6 \pm  6.7 $\\
50 & 1.305 & 1.312 & 1.285 & $ 25.5 \pm  5.5 $ &  1.321 & 1.279 & $ 10.3 \pm 12 $\\
Ave&       &       &       & $ 28.0 \pm  1.9 $ &        &       & $ 21.9 \pm  2.7 $\\
\enddata
\tablecomments{The results of Method II.  Col. (1): ecliptic latitude,
$\beta$. Col. (2) $\csc\beta$ - the value of $\Xi$ in the case of a
uniform slab of dust (see eq. [\ref{equ_xidef}]). Col. (3): $\Xi_K$ -
the value of $\Xi$ at $60\micron$ expected for the Kelsall model and
no background.  Col. (4): Observed value of $\Xi$ at $60\micron$.  Col. (5):
Intensity of background derived from $\Xi_{60}$ (eq. [\ref{equ_bofxi}]).
Col. (6-8): same as col. (3-5) but for $100\micron$.  }
\end{deluxetable}

%----------------------------------------------------------------------
% Table 3
\clearpage
\begin{deluxetable}{l|rr|rrrr}
\tablewidth{0pt}
\tablecaption{Method II Model Dependence
   \label{table_kels}
}
\tablehead{
   \colhead{Par}    &
   \colhead{value}    &
   \colhead{$\sigma$}    &
   \colhead{$\Delta_{60}$}    &
   \colhead{$\Delta_{100}$}    &
   \colhead{$\Delta_{60} (3\sigma)$}    &
   \colhead{$\Delta_{100} (3\sigma)$}    \\
   \colhead{}  &
   \colhead{}  &
   \colhead{}  &
   \colhead{(\%)}  &
   \colhead{(\%)}  &
   \colhead{(\%)}  &
   \colhead{(\%)}  
}
\startdata
  $\Omega$ (deg)    &       77.7 &      0.600 &   0.013 &   0.005 &   0.040 &   0.013 \\
       $i$ (deg)    &       2.03 &     0.0170 &   0.001 &   0.002 &   0.005 &   0.005 \\
  $\alpha$          &       1.34 &     0.0220 &   5.533 &   2.042 &  16.310 &   6.012 \\
  $n_0$ (AU$^{-1}$) &$ 1.13\times10^{-7} $&$ 6.40\times10^{-10} $&   0.020 &   0.009 &   0.058 &   0.027 \\
  $\delta$          &      0.467 &    0.00410 &   1.938 &   0.576 &   5.782 &   1.720 \\
     $T_0$ (K)      &       286. &       5.00 &   1.501 &   0.305 &   4.364 &   0.886 \\
\enddata
\tablecomments{Dependence of IR excess measurements on Kelsall model
       parameters.  The very weak dependence shown here indicates that
either the Kelsall model does not contain the necessary freedom to
adequately describe the zodiacal emission, or else there is a
time-independent component that accounts for the IR excess. 
}
\end{deluxetable}

%----------------------------------------------------------------------
% Table 4
\clearpage
\begin{deluxetable}{l|rrrr}
\tablewidth{0pt}
\tablecaption{Systematic Errors
   \label{table_errors}
}
\tablehead{
   \colhead{Reason}    &
   \colhead{Meth I $\Delta_{60}$}    &
   \colhead{Meth I $\Delta_{100}$}   &
   \colhead{Meth II $\Delta_{60}$}    &
   \colhead{Meth II $\Delta_{100}$}  \\
   \colhead{}  &
   \colhead{$(\nWpMMSr)$}  &
   \colhead{$(\nWpMMSr)$}  &
   \colhead{$(\nWpMMSr)$}  &
   \colhead{$(\nWpMMSr)$}  
}
\startdata
Model           & 5   & 5   & 3  & 1 \\
Algorithmic     & 7   & 11  & 2  & 6 \\
IR Excess Total & 12  & 16  & 5  & 7 \\
%\hline
%WIM         & 6   & 6   & 6  & 6 \\
%CNM         & 2   & 2   & 2  & 2 \\
%Total       & 10  & 12 \\
\enddata
\tablecomments{Summary of systematic uncertainties. 
Model errors reflect the dependence of the IR excess on IPD model
parameters.  Algorithmic errors refer to uncertainty caused by the
details of the implementation of each method.  The sum of these errors
is the systematic uncertainty in the measured IR excess -- errors that
average down when results from the two methods are combined. 
Additional systematic uncertainty results from the WIM and WNM
subtraction; these errors do not average down and are included in
Table \ref{table_results}.
All errors are 95\% confidence. 
}
\end{deluxetable}

%----------------------------------------------------------------------
% Table 5
\clearpage
\begin{deluxetable}{rrr}
\tablewidth{0pt}
\tablecaption{Bad Pixels and Weeks
   \label{table_pixmask}
}
\tablehead{
   \colhead{$\lambda$}    &
   \colhead{Pixels Used}    &
   \colhead{No. Weeks} \\
   \colhead{($\micron$)}  &
   \colhead{\%}  &
   \colhead{}  
}
\startdata
4.9  &  -   &  39  \\
12   &  79  &  39  \\
25   &  80  &  39  \\
60   &  55  &  39  \\
100  &  46  &  31  \\
140  &  82  &  41  \\
240  &  83  &  41  \\
\enddata
\tablecomments{Column (1), DIRBE waveband.  Col. (2) fraction of
pixels used in all good weeks.  Col. (3) Number of good weeks out of
41 science weeks in the DIRBE mission.  The $100\micron$
channel has a large number of weeks discarded due to hysteresis
effects near the north ecliptic pole.  This effect is also present,
but much smaller at $60\micron$. 
}
\end{deluxetable}

%----------------------------------------------------------------------
% Table 6
\clearpage
\begin{deluxetable}{l|rrrr}
\tablewidth{0pt}
\tablecaption{Final results
   \label{table_results}
}
\tablehead{
   \colhead{}    &
   \colhead{$60\micron$}    &
   \colhead{$100\micron$}    &
   \colhead{$140\micron$}    &
   \colhead{$240\micron$}    \\
   \colhead{}  &
   \colhead{$(\nWpMMSr)$}  &
   \colhead{$(\nWpMMSr)$}  &
   \colhead{$(\nWpMMSr)$}  &
   \colhead{$(\nWpMMSr)$}  
}
\startdata
Method I    &$ 29.0 \pm 5.9 \pm 6 $&$ 40.4 \pm 6.5 \pm 8 $&$ 36.6 \pm 19.9 $&$ 14.6 \pm 6.6 $ \\
Method II   &$ 28.0 \pm 1.9 \pm 3 $&$ 21.9 \pm 2.7 \pm 4 $&$      ...      $&$     ...      $ \\
Mean        &$ 28.1 \pm 1.8 \pm 3 $&$ 24.6 \pm 2.5 \pm 4 $&$      ...      $&$     ...      $ \\
WIM/CNM     &$                  4 $&$                  4 $&$      ...      $&$     ...      $ \\
Recommended &$ 28.1 \pm 1.8 \pm 7 $&$ 24.6 \pm 2.5 \pm 8 $&$ 25.0 \pm  6.9 $&$ 13.6 \pm 2.5 $ \\
\enddata
\tablecomments{Summary of results.  In cases where two errors are
shown, the first is statistical and the second is systematic.  All
errors are 1$\sigma$.  Recommended values for $140$ and $240\micron$
are taken from Hauser \etal\ (1998). 
}
\end{deluxetable}
\clearpage

\end{document}